\documentclass[reprint,aps,prb,twocolumn]{revtex4-1}
\usepackage{amsmath,amssymb,mathrsfs}
\usepackage{amsbsy,amsthm,amsfonts}
\usepackage{graphicx}

\allowdisplaybreaks

\begin{document}
\title{Can photonic crystals be homogenized in higher bands?}

\author{Vadim A. Markel}

\thanks{On leave from the Department of Radiology, University of Pennsylvania, Philadelphia, Pennsylvania 19104, USA}
\email{vmarkel@fresnel.fr,vmarkel@mail.med.upenn.edu}
\affiliation{Aix-Marseille Universit\'{e}, CNRS, Centrale Marseille, Institut Fresnel UMR 7249, 13013 Marseille, France}

\author{Igor Tsukerman}

\affiliation{Department of Electrical and Computer Engineering, The University of Akron, OH 44325-3904, USA}
\email{igor@uakron.edu}

\date{\today}

\begin{abstract}
We consider conditions under which photonic crystals (PCs) can be homogenized in the higher photonic bands and, in particular, near the $\Gamma$-point. By homogenization we mean introducing some effective local parameters $\epsilon_{\rm eff}$ and $\mu_{\rm eff}$ that describe reflection, refraction and propagation of electromagnetic waves in the PC adequately. The parameters $\epsilon_{\rm eff}$ and $\mu_{\rm eff}$ can be associated with a hypothetical homogeneous effective medium. In particular, if the PC is homogenizable, the dispersion relations and isofrequency lines in the effective medium and in the PC should coincide to some level of approximation. We can view this requirement as a necessary condition of homogenizability. In the vicinity of a $\Gamma$-point, real isofrequency lines of two-dimensional PCs can be close to mathematical circles, just like in the case of isotropic homogeneous materials. Thus, one may be tempted to conclude that introduction of an effective medium is possible and, at least, the necessary condition of homogenizability holds in this case. We, however, show that this conclusion is incorrect: complex dispersion points must be included into consideration even in the case of strictly non-absorbing materials. By analyzing the complex dispersion relations and the corresponding isofrequency lines, we have found that two-dimensional PCs with $C_4$ and $C_6$ symmetries are not homogenizable in the higher photonic bands. We also draw a distinction between spurious $\Gamma$-point frequencies that are due to Brillouin-zone folding of Bloch bands and ``true'' $\Gamma$-point frequencies that are due to multiple scattering. Understanding of the physically different phenomena that lead to the appearance of spurious and ``true'' $\Gamma$-point frequencies is important for the theory of homogenization.
\end{abstract}

\maketitle

\section{Introduction}
\label{sec:Intro}

The theory of homogenization of periodic electromagnetic media continues to attract significant attention~\cite{hasar_14_1,karamanos_14_1,clausen_14_1,ciattoni_15_1,sozio_15_1} due to the proposed important applications such as sub-wavelength optical imaging~\cite{pendry_00_1}. However, some fundamental questions remain open in this field of research. Perhaps the most important of these questions is the following: to what extent can the ``exotic'' effective parameters obtained via one of the several recently proposed homogenization theories be used without restriction, like the constitutive parameters of homogeneous natural media are conventionally used? Indeed, while it is generally understood that homogenization is an approximate procedure, the accuracy and the applicability range of a given theory is very difficult to ascertain quantitatively. 

In this paper, we investigate whether a photonic crystal (PC) can be homogenized at frequencies above the first bandgap and, particularly, near the $\Gamma$-point. Answering this question is important for the following reason. It is well known that PCs can be characterized by negative dispersion in the higher photonic bands even if the material from which the PC is made has no absorption. On the other hand, in purely dielectric homogeneous media, negative dispersion is obtained only within the absorption bands. A homogeneous material can be simultaneously transparent and characterized by negative dispersion only if it has a nontrivial magnetic permeability~\cite{pokrovsky_02_1,rautian_08_1}. Therefore, to obtain this result by homogenizing an intrinsically non-magnetic PC, one has to consider sufficiently large frequencies where the dispersion is negative~\cite{fn1}. Typically, these frequencies are above the first bandgap of the crystal. We note that some intuitive arguments exist suggesting that a PC can be homogenized sufficiently close to the the $\Gamma$-point frequencies (that is, the frequencies for which the Bloch wave number vanishes). However, a  careful consideration reveals that the first of these frequencies, $\omega_1=0$, is fundamentally different from the higher ones $\omega_n > 0$ ($n = 2, 3, \ldots$). While homogenization can be arbitrarily accurate near $\omega_1$, the same is not true near any $\omega_n$ with $n > 1$.

The above conclusion is consistent with some of the previous numerical investigations of the homogenization problem~\cite{li_06_1,menzel_08_1,menzel_10_2}. In fact, the work reported here is conceptually close to Ref.~\onlinecite{li_06_1}, and one of the main ideas on which the present paper is based has been stated in that reference. Namely, it was noticed that the right-hand side of the dispersion equation $\omega = f({\bf q})$ can be expanded in powers of the Cartesian components of ${\bf q}$ if $\omega$ is close to one of the $\Gamma$-point frequencies $\omega_n$. Here ${\bf q}$ is the Bloch wave vector. Further, for s-polarized waves in two-dimensional PCs with a center of symmetry, this expansion is of the form $f({\bf q}) = \omega_n + \beta_x q_x^2 + \beta_y q_y^2 \ldots$, or in the cases of $C_4$ and $C_6$ symmetries that are considered in this paper, $f({\bf q}) = \omega_n + \beta q^2 \ldots$ Here we have written explicitly the first two non-vanishing terms of the expansion and $\omega$ is assumed to be in the pass-band. By truncating the latter expansion at the second order, we obtain the isotropic isofrequency line $q^2 = (\omega - \omega_n)/\beta$ so that the wave number $q$ does not depend on the direction of propagation. Another observation made in Ref.~\onlinecite{li_06_1} was that, for incident waves with a sufficiently large projection of the wave vector onto the interface (which includes but is not limited to evanescent waves), higher-order terms must be retained in the expansion of $f({\bf q})$ no matter how close the frequency is to a $\Gamma$-point frequency, and that the resultant law of dispersion is no longer isotropic.

Here we develop this basic idea of Ref.~\onlinecite{li_06_1} theoretically and illustrate it with numerical examples of a different kind. While this reference is focused on the transmission and imaging properties of a PC slab, here we consider in detail the dispersion relations and isofrequency surfaces. We work in a 2D geometry and in s-polarization (one-component electric field) so that Maxwell's equations are reduced to a scalar wave equation. Additionally, the medium will have either $C_4$ or $C_6$ symmetry (square or triangular lattice). Under these conditions, only scalar effective parameters can be introduced~\cite{fn1}. The corresponding effective medium is isotropic and so are its isofrequency surfaces. However, we will show that, in the actual PC, the isotropy of isofrequency surfaces is lost in the higher photonic bands once we include waves with sufficiently large projections of the wave vector onto a given axis. Therefore the PC can not be described by local effective parameters.

Thus, we will view two-dimensional PCs with $C_4$ or $C_6$ symmetry whose isofrequency surfaces are not isotropic as not homogenizable~\cite{fn4}. We note that the lack of isotropy can be confused with nonlocality (spatial dispersion). Indeed, nonlocality of material parameters and anisotropy of {\em local} parameters can result in somewhat similar phenomena. But they are not the same effect and should be distinguished. The distinction is especially evident when boundary value problems are considered. For example, scattering from an anisotropic sphere is an analytically-solvable problem of mathematical physics~\cite{stout_06_1,stout_07_1}. However, solving the same problem for a nonlocal sphere is much more complicated and will, generally, yield a different result. 

To complicate things further, it is frequently stated that there exists a complete physical equivalence between two alternative descriptions of the electromagnetic properties of continuous media~\cite{agranovich_book_66,landau_ess_84,agranovich_06_2,agranovich_09_1}. In one description the medium is assigned a nonlocal permittivity $\epsilon(\omega,{\bf q})$ and a trivial permeability $\mu=1$. This is the so-called Landau-Lifshitz approach~\cite{agranovich_09_1}. In the other description, the medium is assigned two local parameters $\epsilon(\omega)$ and $\mu(\omega)$. We have previously argued that the two descriptions are not physically equivalent in general~\cite{markel_13_1}, but it is true that, in the so-called weak nonlocality regime~\cite{fn2}, such an equivalence exists for the refractive index of the medium (but not for the impedance).

For the purpose of this paper, it is unimportant whether the equivalence mentioned above exists or not. If it does exist, then a homogenizable heterogeneous medium should be characterized by some local parameters $\epsilon_{\rm eff}(\omega)$ and $\mu_{\rm eff}(\omega)$. Whether these local parameters correspond to some effective nonlocal parameter $\epsilon_{\rm eff}(\omega,{\bf q})$ and $\mu_{\rm eff} = 1$ in the alternative description is irrelevant. On the other hand, if such local parameters do not exist, then the medium can not be reasonably homogenized and the introduction of the nonlocal permittivity does not solve the problem. Indeed, the knowledge of $\epsilon_{\rm eff}(\omega,{\bf q})$ for all values of its arguments is not sufficient to solve any boundary value problems in a finite sample~\cite{markel_13_1}, unlike the knowledge of the local parameters $\epsilon_{\rm eff}(\omega)$ and $\mu_{\rm eff}(\omega)$. Besides, the typical applications discussed in the literature such as subwavelength imaging require local $\epsilon_{\rm eff}(\omega)$ and $\mu_{\rm eff}(\omega)$. Therefore, we say that, in order for a PC to be homogenizable, its dispersion relation must be (at least, approximately) the same as in a hypothetical homogeneous effective medium with some local parameters $\epsilon_{\rm eff}(\omega)$ and $\mu_{\rm eff}(\omega)$. We emphasize that the above condition is necessary but, in general, not sufficient because it does not include the impedance. But we will show that even this necessary condition of homogenizability does not hold in PCs above the first bandgap.

We illustrate the theoretical arguments of this paper with numerical examples using a rather simple but physically relevant model. As was mentioned above, we consider two-dimensional PCs (hollow inclusions in a high-index host) with a one-component electric field polarized perpendicularly to the plane of periodicity. Such PCs have been previously considered in the literature as homogenizable in the higher photonic bands~\cite{luo_02_1,luo_02_2,craster_11_1}. Similarly to these works, we neglect in the majority of cases frequency dispersion and absorption in the host material. This is done not for computational convenience (our codes can handle the more general case with equal efficiency) but rather to analyze the exact cases that were previously considered in the literature. 

However, to illustrate the effects of absorption, we have also performed simulations for a square-lattice PC with a dispersive and absorptive host. In this case, the higher $\Gamma$-point frequencies can be defined only approximately and the case for homogenizability is even harder to make. Since no qualitatively new phenomena emerge in an absorptive PC, at least, as far as its homogenezability is concerned, we have restricted simulations of triangular-lattice PCs to the case of non-absorbing host.

We finally note that our results can be understood in a more general framework of the {\em uncertainty principle} of homogenization~\cite{tsukerman_16_1}. According to this principle, the larger the deviation of the effective magnetic permeability from unity (according to a given theory), the less accurate this theory is in predicting physical observables such as the transmission and reflection coefficients of a composite slab. 

The remainder of this paper is organized as follows. We start with some general theoretical considerations relevant to the problem at hand in Sec.~\ref{sec:gen} where we explain why the circularity (or sphericity) of a real isofrequency line is not a sufficient condition for homogenizability. Our computational formulas are written down in Sec.~\ref{sec:comp}.
Extensive numerical examples for 2D square- and triangular-lattice PCs are adduced in Sec.~\ref{sec:num}. Sec.~\ref{sec:disc} contains a discussion of the results obtained. The two appendices contain mathematical details pertinent to the law of dispersion in 2D periodic structures. These developments are important for clarification of several subtle points such as classification of the $\Gamma$-point frequencies. Appendix B also gives some rather involved mathematical formulas that can be used in a numerical implementation of the perturbation theory in which the Bloch wave number is viewed as the expansion parameter.

\section{General considerations}
\label{sec:gen}

For Bloch waves with a nonzero fundamental harmonic (as defined in the Appendices) propagating in three-dimensional, intrinsically non-magnetic PCs, the dispersion equation can be written in the following general form~\cite{markel_12_1,markel_13_1} 
\begin{equation}
\label{PC_Disp_Sigma}
\det \left[( {\bf q} \times {\bf q} \times ) + k^2 \Sigma(\omega, {\bf q}) \right] = 0 \ .
\end{equation}
\noindent
Here $k=\omega/c$ is the free-space wave number and $\Sigma(\omega,{\bf q})$ is a 3D tensor, which is completely determined by the PC geometry and composition and by the two arguments $\omega$ and ${\bf q}$. The latter can be considered as mathematically-independent variables and take arbitrary complex values. We will say that a complex vector ${\bf q}$ is the Bloch wave vector of a PC at some frequency $\omega$ if the pair $(\omega, {\bf q})$ satisfies \eqref{PC_Disp_Sigma}. 

Below we work in the frequency domain and restrict the frequency to be real and positive. The set of all ${\bf q}$'s that satisfy \eqref{PC_Disp_Sigma} for a given $\omega > 0$ forms an isofrequency surface in 3D or a line in 2D. However, the words ``surface'' and ``line'' should not be understood literally because ${\bf q}$ is in general complex. For example, in the 2D geometry that we consider below, the set of complex ${\bf q}$'s that satisfy~\eqref{PC_Disp_Sigma} for some $\omega>0$ is a four-dimensional manifold. The isofrequency lines that are commonly displayed are the intersections of this manifold and various two-dimensional sub-spaces.

The function $\Sigma(\omega, {\bf q})$ arises in various physical contexts~\cite{silveirinha_07_1,fietz_10_1,alu_11_2} and is sometimes interpreted as the nonlocal permittivity tensor of the medium due to its explicit dependence on ${\bf q}$. We have shown~\cite{markel_13_1} that the knowledge of $\Sigma(\omega,{\bf q})$ for all values of its arguments is insufficient for solving boundary value problems in finite samples. However, in this paper, we restrict attention to dispersion relations, and to this end the knowledge of $\Sigma(\omega,{\bf q})$ is sufficient. 

In homogeneous nonlocal media, \eqref{PC_Disp_Sigma} is obtained in a straightforward manner by the spatial Fourier transform of the nonlocal susceptibility function. In heterogeneous periodic media, the derivation is more involved. Sometimes \eqref{PC_Disp_Sigma} is derived for such media by employing an external excitation in the form of an ``impressed'' current that overlaps with the medium and has the mathematical form of a plane wave~\cite{silveirinha_07_1,fietz_10_1,alu_11_2}. However, this approach is not necessary and introduction of such currents into the model is not physically justified~\cite{markel_10_2}. Previously, we have derived \eqref{PC_Disp_Sigma} and defined $\Sigma(\omega,{\bf q})$ for general three-dimensional PCs without appealing to the concept of ``impressed'' currents~\cite{markel_12_1,markel_13_1}. Below, in Appendix~\ref{app:A}, we present a basis-free derivation of \eqref{PC_Disp_Sigma} for a 2-dimensional PC. In Appendix~\ref{app:B}, we repeat the derivation using the basis of plane waves, which yields expressions that are directly amenable to numerical computation. 

We can use \eqref{PC_Disp_Sigma} to formally define the $\Gamma$-point frequencies. As one could expect, the definition is confounded by the vector nature of electromagnetic fields. In particular, the $\Gamma$-point frequencies can be polarization-dependent. However, the simulations discussed below have been performed for the special case of transverse Bloch waves, which satisfy the condition ${\bf q} \cdot {\bf E}_0  = 0$, where ${\bf E}_0$ is the amplitude of the fundamental harmonic of the Bloch wave for a given linear polarization state~\cite{markel_12_1}. In this case, \eqref{PC_Disp_Sigma} simplifies to
\begin{equation}
\label{PC_Disp_Scal}
q^2 = k^2 \Sigma(\omega, {\bf q}) \ ,
\end{equation} 
\noindent
where ${\bf q}$ is now a two-dimensional vector orthogonal to ${\bf E}_0$ and $\Sigma(\omega,{\bf q})$ is a scalar (a principal value of the tensor $\Sigma$ that corresponds to the direction of ${\bf E}_0$). Note that ${\bf E}_0$ can be an eigenvector of $\Sigma(\omega,{\bf q})$ for all $\omega$ and ${\bf q}$, typically, as a consequence of the PC symmetry. Also, we use the same notation $\Sigma$ for the tensor and for its principal values, but this should not cause confusion since only the latter interpretation will be used below.  

In what follows, we will consider only the scalar equation \eqref{PC_Disp_Scal} and assume that the two-dimensional Bloch vector ${\bf q} = (q_x, q_y)$ lies in the $XY$ plane of a rectangular frame while ${\bf E}_0 = E_0 \hat{\bf z}$ is collinear with the $Z$-axis~\cite{fn3}. By focusing on the special case of transverse waves, we do not disregard any important effects but rather focus on the essential features of the theory. 

We now proceed with the analysis of Eq.~\eqref{PC_Disp_Scal}. We will say that a frequency $\omega_n$ is the $n$-th $\Gamma$-point frequency of a PC if the following conditions hold:
\begin{subequations}
\label{Gamma_def_gen}
\begin{eqnarray}
\label{Gamma_def_gen_a}
& {\rm (i)}  \  & \omega_n^2 \Sigma(\omega_n,0) = 0 \ ; \\
\label{Gamma_def_gen_b}
& {\rm (ii)} \  & \Sigma(\omega, {\bf q}) \ {\rm is \ an \ analytic \ function \ of \ }  \\
&               & \omega \ {\rm and \ } {\bf q} \ {\rm in \ the \ vicinity \ of} \ \omega=\omega_n, {\bf q}=0 \nonumber
\end{eqnarray}
\end{subequations}
\noindent
The first trivial solution to \eqref{Gamma_def_gen_a} is $\omega_1 = 0$, unless $\Sigma(\omega,0)$ diverges as $1/\omega^2$ or faster at $\omega=0$. This possibility can be safely ignored and the first (fundamental) $\Gamma$-point frequency exists in all PCs, even if they are made of conducting constituents. Also, $\Sigma(0, {\bf q})$ is usually analytic near ${\bf q} = 0$ so the condition \eqref{Gamma_def_gen_b} is also satisfied for $\omega_1$. We do not have a proof of this statement but will see that all counterexamples to \eqref{Gamma_def_gen_b} occur at $\omega_n > 0$.

The higher $\Gamma$-point frequencies can be determined from the equation 
\begin{equation}
\label{Gamma_def_j>1}
\Sigma(\omega_n,0) = 0 \ , \ \ \omega_n > 0 \ .
\end{equation}
\noindent
The condition \eqref{Gamma_def_gen_b} should also hold. It therefore can be seen that the nature of the first and the higher $\Gamma$-point frequencies is different. At the first frequency, $\Sigma(\omega_1,0) = \Sigma(0,0)$ does not turn to zero; in fact, it can be readily seen that $\Sigma(0,0) = n^2_{\rm eff}$, where $n_{\rm eff}$ is the effective refractive index of the medium in the homogenization limit. But for the higher frequencies $\omega_n \neq 0$, we have $\Sigma(\omega_n, 0) = 0$. This equality can not hold in any truly homogeneous medium (see Appendix~\ref{app:A} for more details).

At this point, a few remarks are in order. First, purely real $\Gamma$-point frequencies of order higher than $1$ do not generally exist in PCs with non-negligible absorption. This is so because \eqref{Gamma_def_j>1} is in this case a complex equation: both real and imaginary parts of \eqref{Gamma_def_j>1} must be satisfied simultaneously. This is unlikely to happen at a purely real $\omega$. However, if, as is frequently done in the PC literature, we assume that the dielectric permittivity of the PC constituents is purely real, then \eqref{Gamma_def_j>1} can be expected to have real roots. 

Second, derivation of equations \eqref{PC_Disp_Sigma} or \eqref{PC_Disp_Scal} does not require that ${\rm Re}({\bf q})$ be in the first Brillouin zone (FBZ) of the lattice. However, if ${\bf q}$ is a solution to one of these equations, then ${\bf q} + {\bf g}$ is also a solution, where ${\bf g}$ is any reciprocal lattice vector. Therefore, it is sufficient to consider only the solutions with ${\rm Re}({\bf q}) \in {\rm FBZ}$. 

\begin{figure*}
\includegraphics[width=17cm]{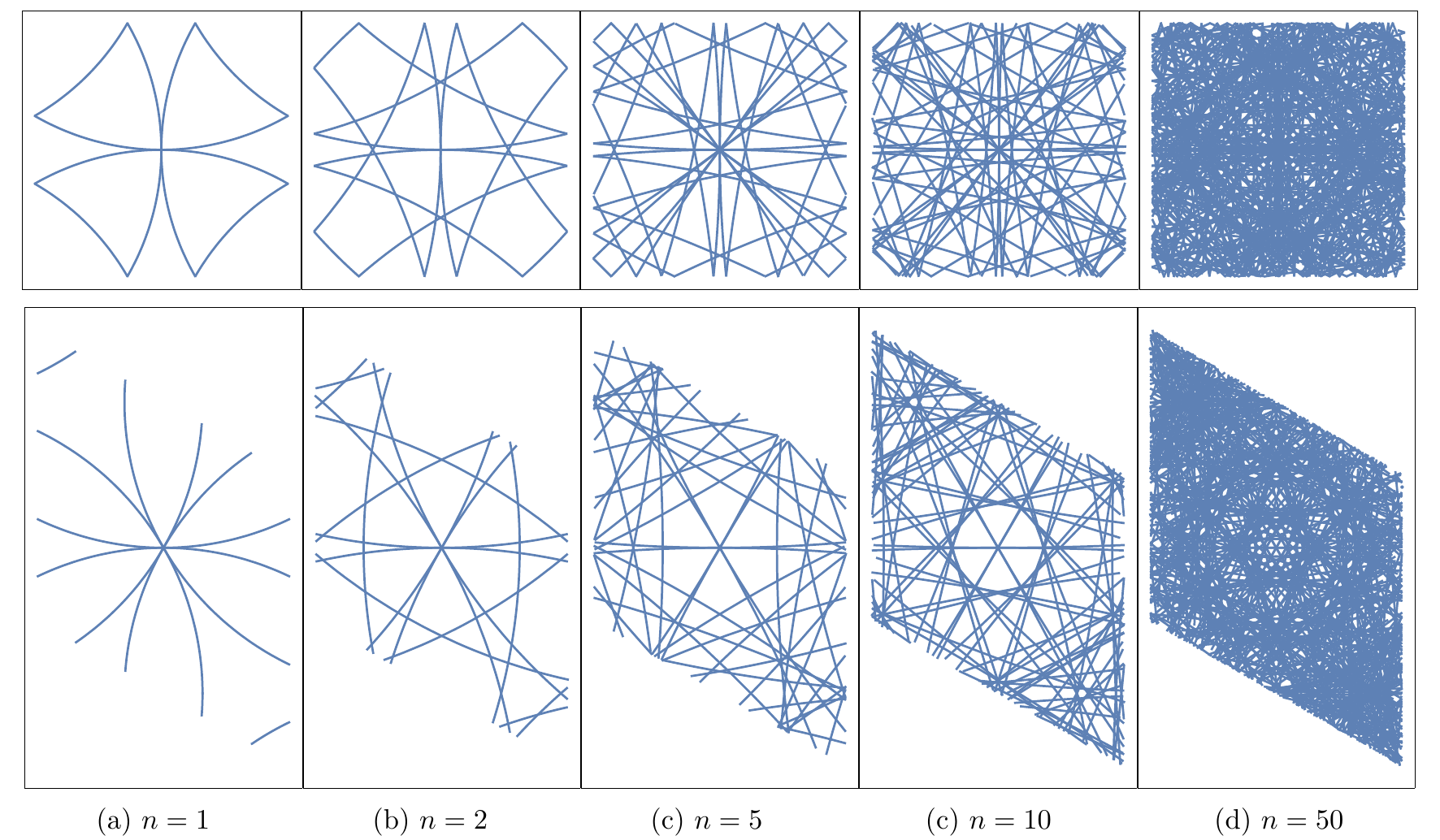}
\caption{(color online) \label{fig:IF_Hmg}
Isofrequency lines for a two-dimensional homogeneous medium artificially discretized on a square (top) and triangular (bottom) lattices. The band index is labeled by $n$. The plots depict the isofrequency lines contained in the FBZ of the lattice and can be periodically replicated in all directions. A mathematical definition of the folding operation is given in Eq.~\eqref{IFE_HMG_PER}.}
\end{figure*}

Third, and related to the above, homogeneous media do not possess higher-order $\Gamma$-point frequencies in the sense of definition \eqref{Gamma_def_gen}. It is true that the ``artificially folded'' dispersion curve of a homogeneous medium, such as the one shown in figure 2 of Ref.~\onlinecite{joannopoulos_book_08}, crosses the vertical axis at some frequencies $\omega_n$, and one can conclude that at $\omega = \omega_n$ we also have ${\bf q} = 0$. However, condition \eqref{Gamma_def_gen_b} is not satisfied at these frequencies. This point is illustrated in Fig.~\ref{fig:IF_Hmg} where we plot the purely real isofrequency lines for a homogeneous medium that is artificially discretized on two-dimensional square and triangular lattices. It can be seen that the isofrequency lines are very far from circular and become quasi-chaotic for large band indexes $n$. This behavior distinguishes these pseudo-$\Gamma$-point frequencies from the true ones, which  can exist in PCs due to strong multiple scattering. For a more detailed analysis, see Appendix~\ref{app:A}.

Returning to the question of homogenizability, we can now see why one might be tempted think that the medium to be homogenizable in the vicinity of a true $\Gamma$-point frequency. Let $\omega$ be in the transmission band of a PC but close to a $\Gamma$-point frequency $\omega_n$ ($n > 1$), say, slightly below $\omega_n$. Let us fix the frequency and expand $\Sigma(\omega, {\bf q})$ in powers of ${\bf q}$. This is possible because of \eqref{Gamma_def_gen_b}. If, as is often the case, the PC has a center of symmetry, the expansion is to lowest order in ${\bf q}$ of the form
\begin{equation}
\label{sigma_exp_2}
\Sigma(\omega, {\bf q}) = \alpha + \beta_x \frac{q_x^2}{k^2} + \beta_y \frac{q_y^2}{k^2} + \ldots \ , \ \ \alpha \equiv \Sigma(\omega,0)\ll 1 \ .
\end{equation}
\noindent
In the above equation, $\omega$ is a fixed parameter and therefore the explicit dependence of the expansion coefficients $\alpha$, $\beta$ on $\omega$ is suppressed. Also, it is easy to see that in PCs obeying $C_4$ or $C_6$ symmetry, $\beta_x = \beta_y = \beta$. If we now treat the first two terms in the expansion \eqref{sigma_exp_2} as an approximation, write $\Sigma(\omega,{\bf q}) \approx \alpha + \beta q^2/k^2$ and substitute the result into the dispersion equation \eqref{PC_Disp_Scal}, we will obtain an approximate solution of the form
\begin{equation}
\label{q_2}
q^2 = k^2 \frac{\alpha}{1 - \beta} \ .
\end{equation}
\noindent
We remark briefly that it is rather typical to interpret $\alpha$ as the effective permittivity of the medium, $\epsilon_{\rm eff}$, and $1/(1 - \beta)$ as the effective permeability, $\mu_{\rm eff}$. This approach is discussed in detail in Ref.~\onlinecite{markel_13_1}. Since we consider only dispersion relations in infinite media in this paper (a connection to the problem of transmission and reflection of waves is made by observing that the tangential component of the Bloch wave vector is conserved at planar interfaces), the breakdown of the squared refractive index $n^2_{\rm eff}$ into the product of $\epsilon_{\rm eff}$ and $\mu_{\rm eff}$ is irrelevant. What is important for our purposes is that \eqref{q_2} describes a perfectly circular isofrequency line. Of course, circular isofrequency lines are also characteristic of homogeneous materials. The conclusion is then drawn that sufficiently close to a $\Gamma$-point frequency, a PC is indistinguishable from a homogeneous medium.

However, the above line of argument has the following deficiency. It is not really true that all Cartesian components of ${\bf q}$ must be small near a higher-order $\Gamma$-point frequency. A more precise statement is that the scalar $q^2 = {\bf q} \cdot {\bf q}$ is small. The Cartesian components of ${\bf q}$ can still be arbitrarily large for complex ${\bf q}$. An obvious example is the vector ${\bf q} = (p,ip)$ where $p$ is a real number. 

In Appendix~\ref{app:B}, we develop a perturbation theory for $\Sigma(\omega,{\bf q})$. The expansion is obtained for two-dimensional PCs with either $C_4$ or $C_6$ symmetry in terms of the Cartesian components of ${\bf q}$ for $\omega$ in the vicinity of one of the $\Gamma$-point frequencies $\omega_n$ as defined in \eqref{Gamma_def_gen}. The result is a special case of the more general expansion \eqref{sigma_exp_2} (which does not assume any special symmetry) and is of the form
\begin{eqnarray}
\label{Sigma_beta}
\Sigma(\omega,{\bf q}) &= \alpha + \beta_2 q^2 + \beta_4 q^4 + \beta_6 q^6 + \beta_8 q^8 + \gamma_4 q_x^2 q_y^2 \nonumber \\
&+ \left[ 
 \delta_{6x} q_x^2 \left(q_x^4 - 6q_x^2 q_y^2 + 9 q_y^4 \right) \right. \nonumber \\
&+ \left.
 \delta_{6y}q_y^2 \left(q_y^4 - 6q_x^2 q_y^2 + 9 q_x^4 \right)
\right] + \ldots
\end{eqnarray}
\noindent
Here $q^4 = (q_x^2 + q_y^2)^2$, etc. It can be seen that the expansion terms with the coefficients $\beta_{2n}$ are all circularly-symmetric. The first term that breaks the circular symmetry is
\begin{equation}
\label{C4}
q_x^2 q_y^2 \ .
\end{equation} 
\noindent
We can say that, starting from fourth order in ${\bf q}$, the function $\Sigma(\omega, {\bf q})$ starts to bear the traces of the underlying lattice, which is not circularly-symmetric. 

However, the coefficient $\gamma_4$ is zero in the case of $C_6$ symmetry of the structure. This is so because expression \eqref{C4} is not invariant with respect to rotation by $\pi/3$. The simplest anisotropic terms that are invariants of $C_6$ arise in sixth order in ${\bf q}$ and are of the form
\begin{subequations}
\label{C6}
\begin{eqnarray}
\label{C6a}
q_x^2 \left(q_x^4 - 6q_x^2 q_y^2 + 9 q_y^4 \right) \ , \\
\label{C6b}
q_y^2 \left(q_y^4 - 6q_x^2 q_y^2 + 9 q_x^4 \right) \ .
\end{eqnarray}
\end{subequations}
\noindent
The coefficients $\delta_{6x}$ and $\delta_{6y}$ in front of these terms in \eqref{Sigma_beta} are rather complicated and we have not computed them explicitly in Appendix~\ref{app:B}. We note, however, that the $X$ and $Y$-directions are not equivalent in a triangular lattice. Therefore, $\delta_{6x} \neq \delta_{6y}$. If these two coefficients were equal, the term in the square brackets in \eqref{Sigma_beta} would reduce to $q^6$. This is exactly what happens in the case of a square lattice where the $X$ and $Y$-directions are equivalent and the corresponding coefficients are absorbed in $\beta_6$.

\begin{figure}
\centerline{\includegraphics[width=8.2cm]{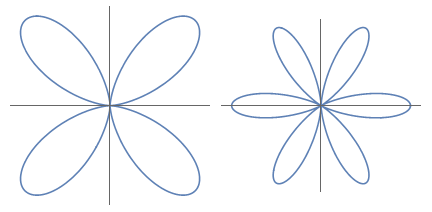}}
\centerline{(a) Term \eqref{C4} \hspace{1.8cm} (b) Term \eqref{C6a}}
\caption{\label{fig:IF_C4C6}
Graphical illustration of the terms \eqref{C4} and \eqref{C6a}. The parametric plots show the dependence of the magnitude of each term on the direction of the (purely real) vector ${\bf q}$. Note that the lines shown in plots (a) and (b) are similar (but not completely identical) to the isofrequency lines shown in Fig.~\ref{fig:IF_Hmg} for $n=1$. In the latter case, the lines are reflected by the boundaries of the FBZ. However, in the vicinity of the origin, the lines are identical [up to a $\pi/6$ rotation for the term \eqref{C6a}]. This is a general manifestation of the applicable rotational symmetry group [$C_4$ in (a) and $C_6$ in (b)].} 
\end{figure}

The terms \eqref{C4} and \eqref{C6a} are graphically illustrated in Fig.~\ref{fig:IF_C4C6}. 

We thus see why the triangular lattice is more amenable to homogenization: the anisotropic terms start to appear in this case only in sixth order in ${\bf q}$. However, as soon as these terms yield a noticeable contribution to $\Sigma(\omega,{\bf q})$, isotropy is lost very fast. This observation was made in Ref.~\onlinecite{li_06_1} and illustrated by considering the transmission coefficient of a PC slab. Below we will illustrate this observation by plotting the isofrequency lines for an extended range of $q_x$, which includes not only propagating incident waves (in a vacuum or in a PC or in both), but also evanescent incident waves. 

In what follows, we will consider the following problem. Fix the frequency and assume that the Bloch wave vector has a known and purely real projection $q_x$ onto an axis that is tangential to the PC/air interface (although the interface is not considered explicitly). Then compute the corresponding values of $q_y$ that satisfy the dispersion equation. Here $q_y$ can be real, imaginary or complex, even if the PC is made of lossless components. In a lossless PC, the set of purely real solutions $(q_x,q_y)$ would form a traditional isofrequency line. However, the dispersion equation has a solution (in fact, infinitely many solutions) for any $q_x$. Some of these solutions are complex. When $q_y$ has a nonzero imaginary part, the corresponding Bloch wave is evanescent. Evanescent Bloch waves can be excited in the PC by an incident plane wave that is either propagating or evanescent in a vacuum. It is important to note, however, that the projection of the incident wave vector onto any flat interface is equal to $q_x$. Thus, to make sure that a homogenization theory is applicable to a sufficiently large class of  incident waves, which must necessarily include evanescent waves.

\section{Computational formulas and numerical methods}
\label{sec:comp}

In the numerical examples we consider a two-dimensional PC whose exact permittivity $\tilde{\epsilon}({\bf r})$ satisfies the periodicity relation
\begin{equation}
\label{latt_per}
\tilde{\epsilon}({\bf r} + n_1 {\bf a}_1 + n_2 {\bf a}_2) = \tilde{\epsilon}({\bf r}) \ ,
\end{equation}
\noindent
where ${\bf a}_1$ and ${\bf a}_2$ are two primitive lattice vectors and $n_1$, $n_2$ are integers. Here and below the overhead tilde will be used to denote functions obeying the lattice periodicity \eqref{latt_per}. We will work in an orthogonal reference frame whose $Z$-axis is perpendicular to both ${\bf a}_1$ and ${\bf a}_2$. Then these two vectors lie in the $XY$ plane. We will further consider electromagnetic waves with one-component electric field $\hat{\bf E} = (0,0,E)$ and two-component magnetic field ${\bf H}=(H_x,H_y,0)$. In this case, Maxwell's equations are reduced to the scalar wave equation for the electric field
\begin{equation}
\label{wave_eq}
\left [ \nabla^2 + k^2 \tilde{\epsilon}({\bf r}) \right] E({\bf r}) = 0 \ ,
\end{equation}
\noindent
where ${\bf r} = (x,y)$. 

The formulas of this section are applicable to arbitrary primitive vectors ${\bf a}_1$ and ${\bf a}_2$. However, the simulations will be performed for the square lattice ($a_1 = a_2 = a$ and ${\bf a}_1 \cdot {\bf a}_2 = 0$) and equilateral triangular lattice ($a_1 = a_2 = a$ and ${\bf a}_1 \cdot {\bf a}_2 = a^2/2$).

Let ${\bf b}_1$ and ${\bf b}_2$ be the primitive vectors of the reciprocal lattice such that
\begin{equation}
{\bf a}_j \cdot {\bf b}_k = 2\pi \delta_{jk} \ .
\end{equation}
\noindent
A generic reciprocal lattice vector ${\bf g}$ can be written as
\begin{equation}
{\bf g} = {\bf b}_1 n_1 + {\bf b}_2 n_2 \ , 
\end{equation}
\noindent
where $n_1, n_2$ are integers. We can view ${\bf g}$ as a discrete composite index, which maps one-to-one to the pair $(n_1,n_2)$. In particular, the exact permittivity of the medium is expandable as
\begin{equation}
\label{eps_exp}
\tilde{\epsilon}({\bf r}) = \sum_{\bf g} \epsilon_{\bf g} e^{i {\bf g} \cdot {\bf r}} \ , \ \ \epsilon_{\bf g} = \frac{1}{S[{\mathbb C}]} \int_{\mathbb C} \tilde{\epsilon}({\bf r}) e^{-i {\bf g} \cdot {\bf r}} d^2 r \ ,
\end{equation}
\noindent
where ${\mathbb C}$ is an elementary cell of the medium and $S[{\mathbb C}]$ is its area, i.e., $a^2$ for a square lattice or $\vert ({\bf a}_1 \times {\bf a}_2) \cdot \hat{\bf z} \vert$ in the more general case. 

Our simulations were performed for a two-component PC and we therefore adduce the computational formulas that are specific to this case. However, we believe that the conclusions of this paper are not limited to two-component media. For a two-component medium we can write
\begin{equation}
\label{eps_two_comp}
\tilde{\epsilon}({\bf r}) = \epsilon_h + (\epsilon_i - \epsilon_h) \tilde{\Theta}({\bf r}) \ , 
\end{equation}
\noindent
where $\epsilon_h$ and $\epsilon_i$ are the permittivities of the host and the inclusions and $\tilde{\Theta}({\bf r})$ is the lattice-periodic shape function. Let the region of the inclusion be $\Omega \in \mathbb{C}$. Then $\tilde{\Theta}({\bf r})  = 1$ if ${\bf r} \in \Omega$ and $\tilde{\Theta}({\bf r}) = 0$ otherwise. Upon Fourier transformation, we find that
\begin{equation}
\label{eps_g}
\epsilon_{\bf g} = \epsilon_h \delta_{{\bf g}0} + \rho\chi M({\bf g}) \ ,
\end{equation}
\noindent
where
\begin{subequations}
\label{M_rho_chi_def}
\begin{align}
& M({\bf g}) = \frac{1}{S[\Omega]}\int_\Omega e^{-i {\bf g} \cdot {\bf r}} d^2 r \ , \\
& \rho = S[\Omega]/S[\mathbb{C}] \ , \ \ 
  \chi = \epsilon_i - \epsilon_h \ .
\end{align}
\end{subequations}
\noindent
In the above equations, $\rho$ is the area fraction of the inclusions and $\chi$ is the contrast. The function $M({\bf g})$ contains information about the inclusion geometry but is independent of $\epsilon_h$ and $\epsilon_i$.

Bloch-periodic functions such as the electric field $E({\bf r})$ and displacement $D({\bf r})$ can be expanded as
\begin{equation}
\label{Bloch_exp}
E({\bf r}) = \sum_{\bf g} E_{\bf g} e^{i ({\bf q} + {\bf g}) \cdot {\bf r}} \ , \ 
D({\bf r}) = \sum_{\bf g} D_{\bf g} e^{i ({\bf q} + {\bf g}) \cdot {\bf r}} \ ,
\end{equation}
\noindent
where ${\bf q}$ is the Bloch wave vector, which must be determined by substituting \eqref{Bloch_exp} into \eqref{wave_eq}. The equation $D({\bf r}) = \tilde{\epsilon}({\bf r})E({\bf r})$ takes the form $D_{\bf g} = \sum_{\bf p} \epsilon_{{\bf g} - {\bf p}} E_{\bf p}$
while \eqref{wave_eq} is reduced in this case to
\begin{equation}
\label{E_D_Maxwell}
({\bf g} + {\bf q})^2 E_{\bf g} = k^2 D_{\bf g} \ .
\end{equation}
\noindent
Combining the wave equation and constitutive relation together we obtain the eigenproblem
\begin{equation}
\label{main_gen}
({\bf g} + {\bf q})^2 E_{\bf g} = k^2 \sum_{\bf p} \epsilon_{{\bf g}-{\bf p}} E_{\bf p} \ ,
\end{equation}
\noindent
which determines the allowable values of the vector ${\bf q}$ for each frequency $\omega$ or for the corresponding wavenumber $k = \omega/c$. We now use the expression for $\epsilon_{\bf g}$ \eqref{eps_g}, which is specific to two-component PCs.  Upon substitution of \eqref{eps_g} into \eqref{main_gen}, we obtain the infinite set of equations
\begin{equation}
\label{main}
\left[ ({\bf g} + {\bf q})^2 - k^2 \epsilon_h \right] E_{\bf g} = \rho \chi k^2 \sum_{\bf p} M({\bf g}-{\bf p}) E_{\bf p} \ .
\end{equation}
\noindent
For each real value of $q_x$ (and $k>0$), the above equation defines an infinite set of $q_y$. 

In the simulations, we have solved \eqref{main} by linearization and truncation of the grid of ${\bf g}'s$. Linearization can be achieved by the standard method of defining the new variable ${\bf H}_{\bf g} = ({\bf g} + {\bf q})E_{\bf g}$. For any finite truncation of the grid of ${\bf g}$'s, the operation of linearization increases the number of equations by the factor of $3$. The resultant linear eigenproblem is neither Hermitian nor symmetric and, therefore, its solutions are in general complex. 

We have solved the above eigenproblem by using Intel's MKL library of subroutines for Fortran. Convergence of the results with the grid size was checked by (i) consecutively doubling the size of the grid and (ii) by comparison with a high-order finite difference method (FLAME)~\cite{tsukerman_08_2}. We have obtained both an excellent convergence with respect to the grid size and an excellent agreement with the finite-difference method, which allows us to conclude that the numerical results shown below are accurate.

Below, we compare the solutions obtained numerically for the actual PC to similar solutions in a homogeneous effective medium, which is artificially discretized on the same lattice as that of the PC. The refractive index $n_{\rm eff}$ of the medium is determined from the approximate radius of the purely real quasi-circular lobe of the isofrequency line of the PC (or from a more general fit in the case of losses). This way, the effective medium mimics the dispersion law of the PC as accurately as possible for the directions of propagation that correspond to the points on the quasi-circular lobe. To perform the comparison for more general Bloch wave vectors ${\bf q}$, we need to introduce the operation of ``folding'' into the FBZ of the lattice. This operation can be formally defined as
\begin{subequations}
\label{IFE_HMG_PER}
\begin{align}
\label{q_FBZ_q}
& \left[ {\bf q} \right]_{\rm FBZ} = {\bf q} - \left( n_1 {\bf b}_1 + n_2 {\bf b}_2 \right) \ , \\
& n_i = {\rm Nint}\left( \frac{{\bf q} \cdot {\bf a}_i}{2\pi} \right) \ .
\end{align}
\end{subequations} 
\noindent
In the above equations, ${\rm Nint}(z)$ is the nearest integer to the complex number $z$. The vector ${\bf q}$ in the right-hand side of \eqref{q_FBZ_q} is not restricted to the FBZ and, in the effective medium, it satisfies the dispersion equation $q_x^2 + q_y^2 = n_{\rm eff}^2 k^2$ where $n_{\rm eff}$ is the effective index of refraction. We can view $q_x$ as a mathematically-independent real-valued variable, compute $q_y$ as $q_y = \pm \sqrt{n_{\rm eff}^2 k^2 - q_x^2}$, substitute the resultant pair $(q_x,q_y)$ in the right-hand side of \eqref{q_FBZ_q}, and this will yield the dispersion equation of the artificially discretized homogeneous effective medium. Examples of such folding will be shown in the figures below, and Fig.~\ref{fig:IF_C4C6} was obtained by plotting all real-valued points $\left[ {\bf q} \right]_{\rm FBZ}$ in the plane $(q_x,q_y)$ at the particular frequencies for which these lines cross the origin. 

\section{Numerical examples}
\label{sec:num}

\subsection{Square lattice with non-dispersive and non-absorbing host}
\label{subsec:square}

Consider a two-dimensional square lattice of infinite hollow cylinders embedded in a high-index host. The primitive lattice vectors are in this case ${\bf a}_1 = a(1,0)$, ${\bf a}_2 = a(0,1)$ and for the reciprocal lattice, ${\bf b}_1 = \frac{2\pi}{a}(0,1)$ and ${\bf b}_2 = \frac{2\pi}{a}(1,0)$. The radius of the cylinders is taken to be $R=0.33a$. The host permittivity is $\epsilon_h = 9.61$ and the inclusions (cylinders) are assumed to be a vacuum with $\epsilon_i = 1$. In this example, we disregard absorption either in the host or in the inclusions. This is an ``inversion'' of the model used in Ref.~\onlinecite{gajic_05_1} where a 2D square lattice of aluminum oxide rods in air was considered.

We start with the purely real dispersion diagram. The latter is obtained by setting $q_x=0$, computing $q_y$ for a range of electromagnetic frequencies, and by keeping only real solutions $q_y$ to the dispersion equation. That is, we will disregard for the moment all complex and imaginary wave numbers $q_y$ (these solutions will be considered later). This approach is conventional for purely real permittivities of the constituent material. Note that we compute the dispersion diagram for a given propagation direction (along the $Y$-axis). Dispersion curves for different propagation directions are, generally, different. However, if a higher $\Gamma$-point frequency $\omega_n$ exists according to the definition \eqref{Gamma_def_j>1}, then $\omega_n$ is independent of the propagation direction.

\begin{figure}
\includegraphics[width=8.2cm]{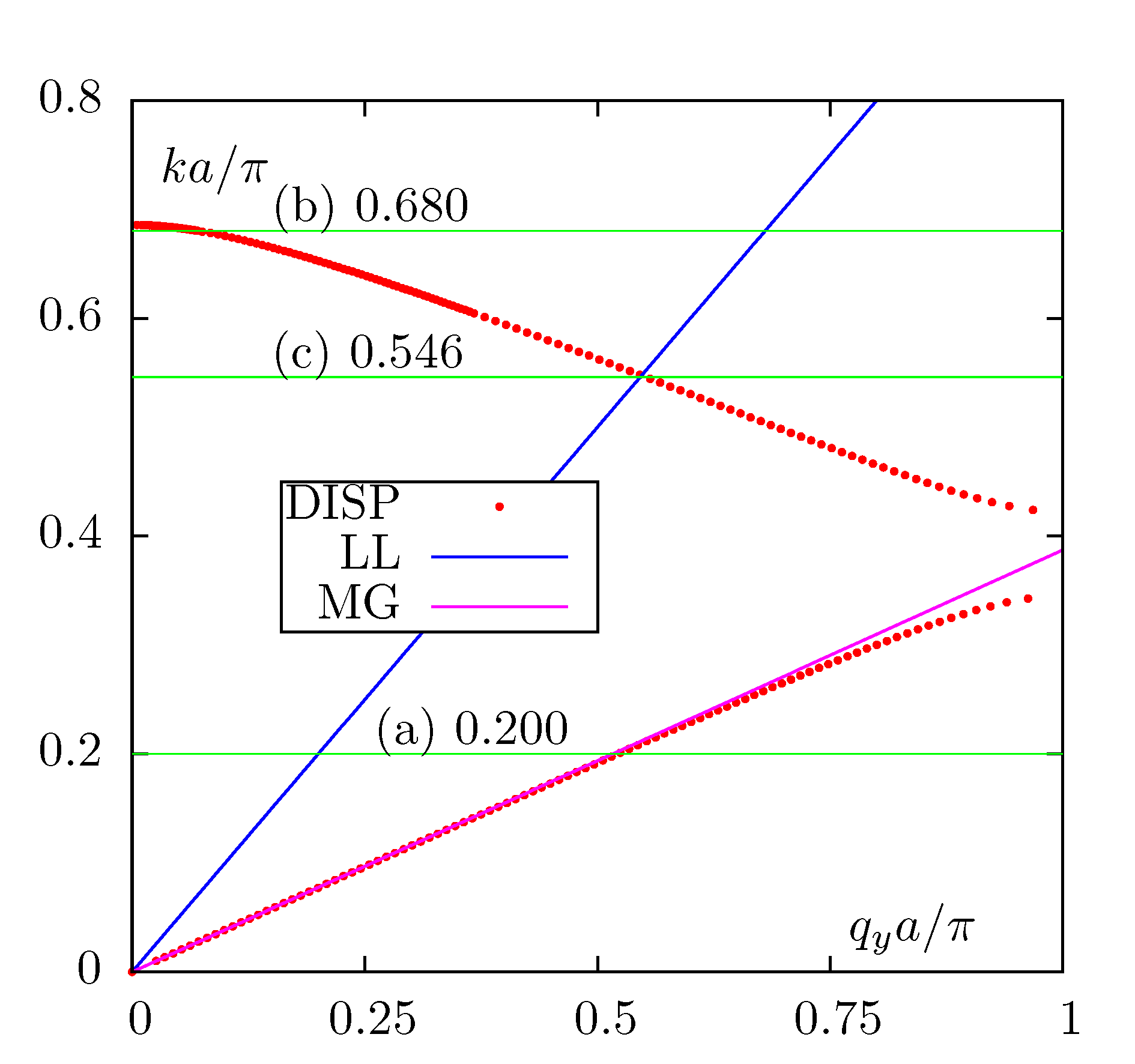}
\caption{(color online) \label{fig:Disp_Circ}
  Dispersion diagram for the square lattice of hollow cylinders. Frequency is scanned from zero to slightly above the second $\Gamma$-point frequency. The variable $ka/\pi = \omega a / \pi c = 2 a / \lambda$ is the dimensionless frequency. The line labeled LL is the light line and the line labeled MG gives the Maxwell-Garnett approximation to the dispersion curve. Thin horizontal lines mark the frequencies at which further simulations have been performed.}
\end{figure}

The dispersion diagram for the square lattice of hollow cylinders is shown in Fig.~\ref{fig:Disp_Circ}. The horizontal lines in this figure mark the dimensionless frequencies at which further simulations have been performed. 

The frequency $ka/\pi=0.200$ [Case (a)] is in the first photonic band. The squared effective refractive index at this frequency is $n^2_{\rm eff} = 6.81$. Note that the Maxwell-Garnett homogenization result for Case (a) is slightly different: $n^2_{\rm eff} = 6.66$. 

The frequency $ka/\pi=0.680$ [Case (b)] is in the second transmission band, very close to and slightly below the second $\Gamma$-point frequency. This is the main case we are interested in. 

Finally, the frequency $ka/\pi=0.546$ [Case (c)] is at the intersection of the second dispersion branch and the light line. One can expect $n_{\rm eff}^2 \approx 1$ and $n_{\rm eff}\approx -1$ at this frequency, since the dispersion in the second photonic band is negative. However, we will see that the medium is very far from being homogenizable in cases (b) and (c) and can not be characterized by local effective parameters.

\begin{figure*}
\includegraphics[width=16.4cm]{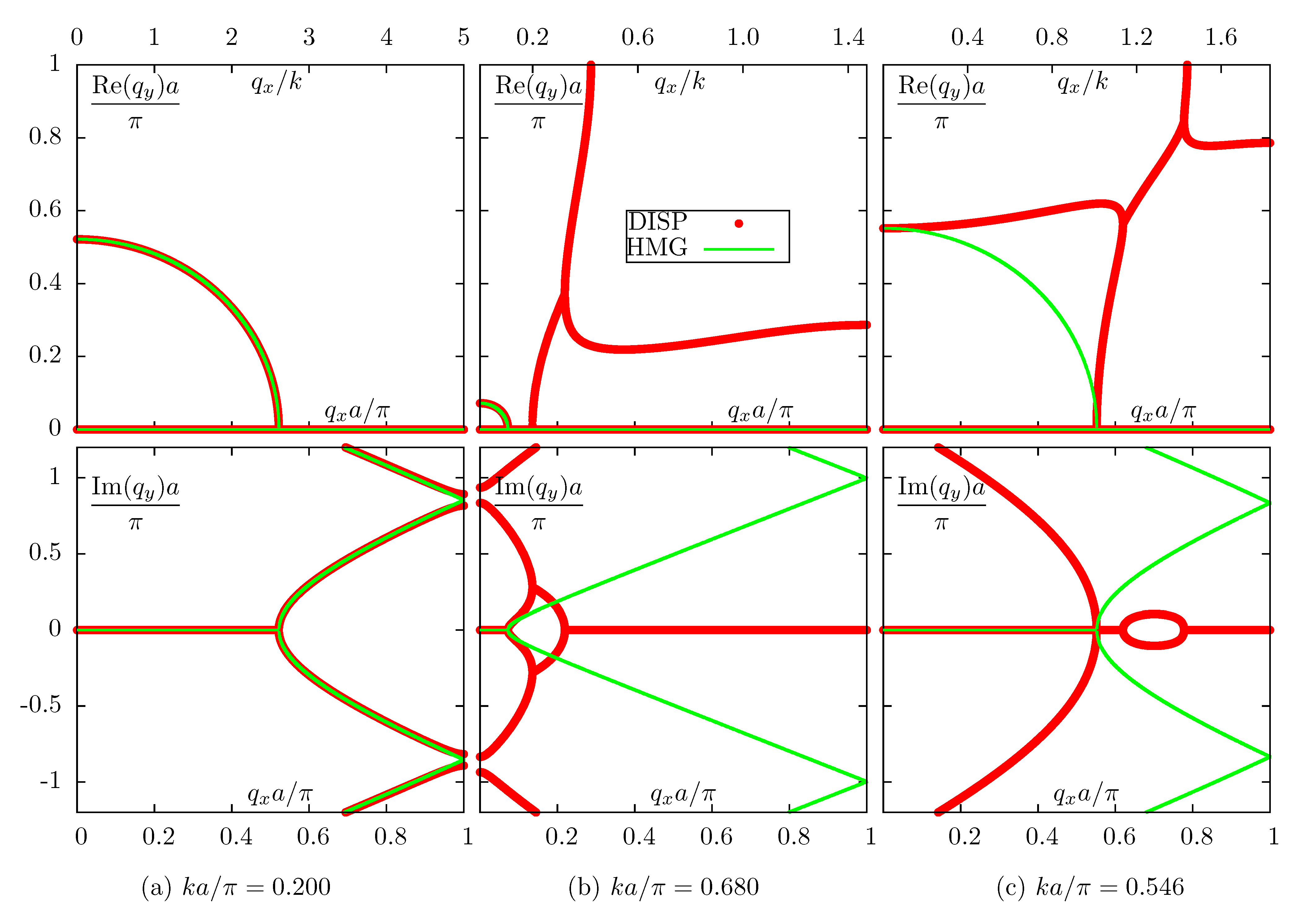}
\caption{(color online) \label{fig:IF_Circ}
  Real (upper row of plots) and imaginary (bottom row) parts of $q_y$ as functions of $q_x$ (real by definition) for the square lattice of hollow cylinders at various values of the dimensionless frequency $ka/\pi$. Dielectric permittivities of the host and inclusions are $\epsilon_h=9.61$, $\epsilon_i=1$; the ratio of the cylinder radius to the lattice period is $R/a=0.33$.Only one quadrant of the FBZ is shown in the upper row of plots; this quadrant can be replicated to cover the whole FBZ. The scale of the lower horizontal axes is $\pi/a$ and the scale of the upper horizontal axes is $1/k$, as labeled. All data points with ${\rm Re}(q_y) = 0$ (shown in the upper row) have nonzero imaginary parts except for the data point $q_y=0$ at the intersection of the quasi-circular lobe and the horizontal axis. The corresponding values of $q_y$ are purely imaginary. The imaginary parts of some of these data points are outside of the plotting range in the lower row of plots. Data points are labeled 'DISP' for the actual PC  and 'HMG' for a homogeneous medium that is artificially discretized on the same lattice. The HMG isofrequency lines were computed according to \eqref{IFE_HMG_PER} with $n_{\rm eff}^2 =  6.81$ (a),  $n_{\rm eff}^2 = 0.011$ (b) and $n_{\rm eff}^2 = 0.91$ (c). In the case (c), the choice of $n_{\rm eff}^2$ guarantees the correct wave number for propagation along $X$- or $Y$-axes but not for the intermediate directions.}
\end{figure*}

\begin{figure*}
\includegraphics[width=16.4cm]{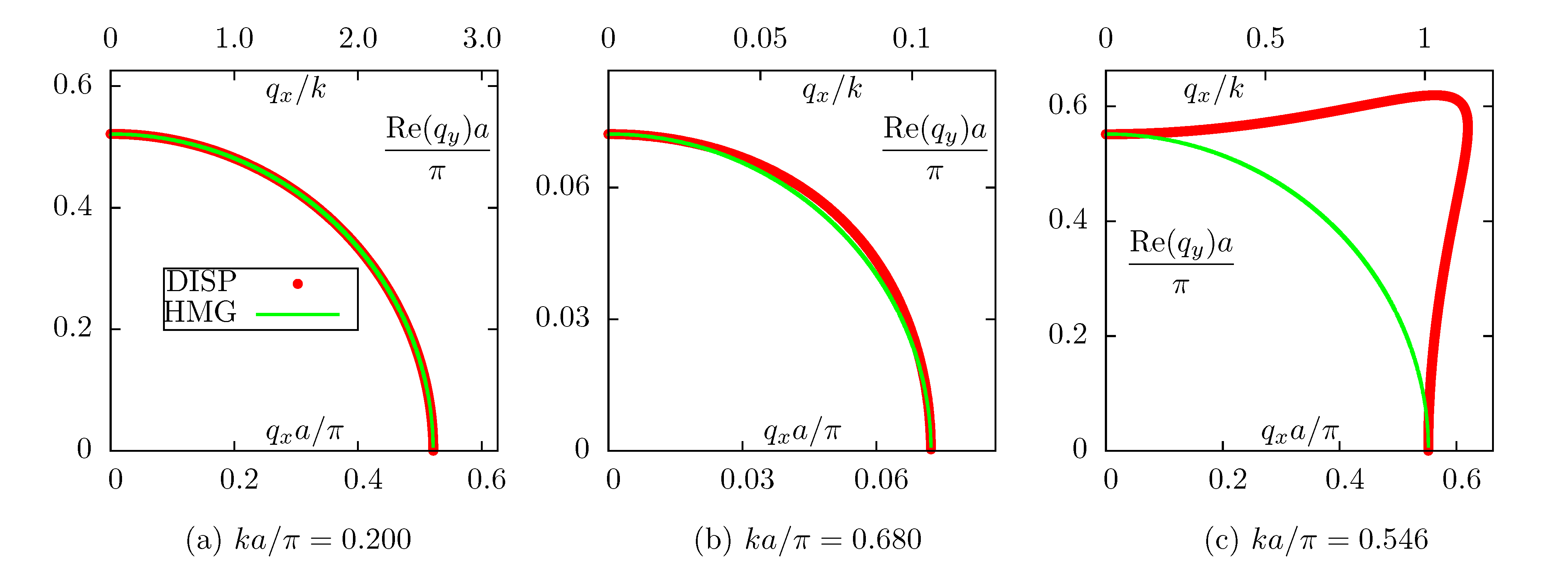}
\caption{(color online) \label{fig:IF_Circ_Arcs}
Purely real parts of the isofrequency lines shown in Fig.~\ref{fig:IF_Circ}. The lines marked 'HMG' are ideal circles.}
\end{figure*}

\begin{figure*}
\includegraphics[width=16.4cm]{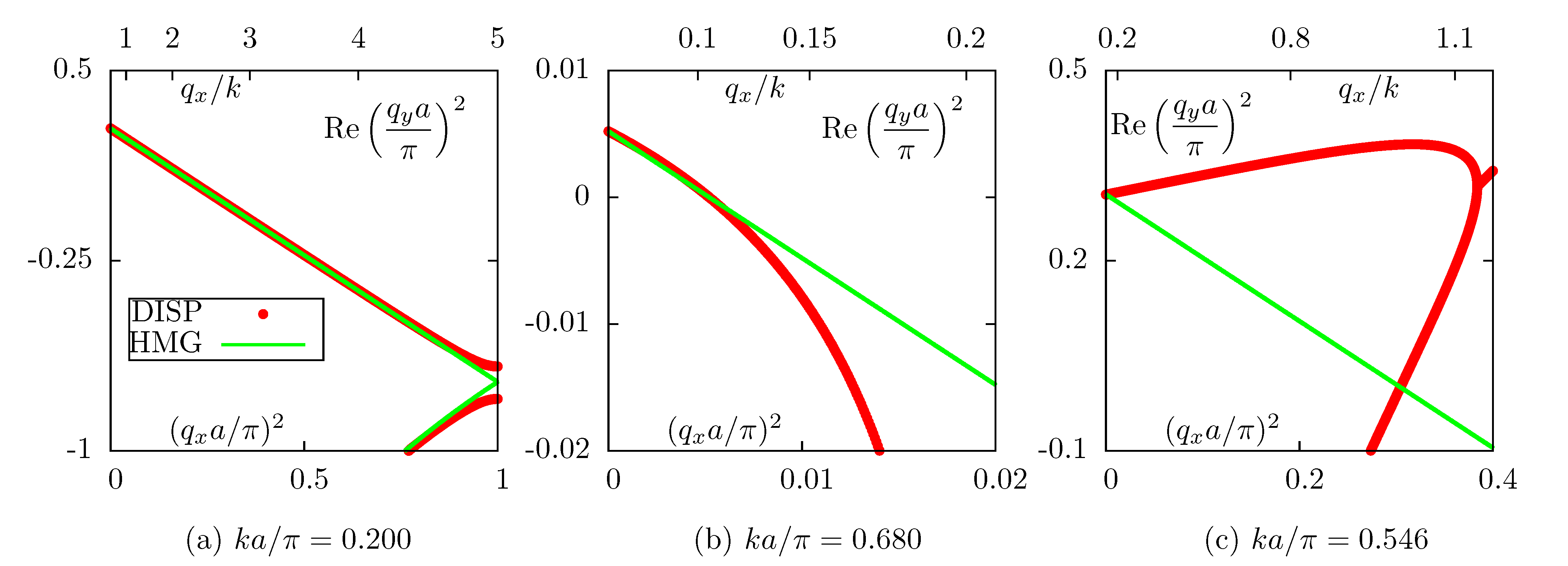}
\caption{(color online) \label{fig:IF_Circ_Squares} Isofrequency lines for the squares of the Cartesian components of Bloch wave vector ${\bf q}$. More precisely, ${\rm Re}(q_y^2)$ is plotted as a function of $q_x^2$ for the same set of fixed frequencies that were considered in Figs.~\ref{fig:IF_Circ} and \ref{fig:IF_Circ_Arcs}. In an effective homogeneous medium artificially discretized on a square lattice, this dependence is piece-wise linear as is shown in panel (a).}
\end{figure*}

The isofrequency lines for the three frequencies noted above are shown in Figs.~\ref{fig:IF_Circ}-\ref{fig:IF_Circ_Squares}. As was explained in Sec.~\ref{sec:gen}, we fix the frequency and view $q_x$ as a mathematically-independent and purely real variable, which is scanned in some interval. For each $q_x$ considered, we find all values of $q_y$ whose imaginary parts are restricted to some sufficiently large range.

In Fig.~\ref{fig:IF_Circ}, we plot the real and imaginary parts of $q_y$ as functions of $q_x$. For each $q_x$, there exist infinitely many solutions to the dispersion equation with ${\rm Re}(q_y) = 0$ and ${\rm Im}(q_y) \neq 0$ and we can not display all such points in the plots. However, the number of solutions with ${\rm Re}(q_y) \neq 0$ is finite, and all such data points are shown in the figure. 

Note that, for each solution $(q_x,q_y)$ in which both $q_x$ and $q_y$ are real, there is also a solution $(q_y,q_x)$. However, this symmetry is broken if $q_y$ has a nonzero imaginary part. For this reason, the upper plots are not completely symmetric with respect to the line $q_y=q_x$. However, the lobes in the lower-left and upper-right corners consist of purely real solutions and these lobes are symmetric. The line that connects these two lobes consists of complex solutions $q_y$ with both real and imaginary parts different from zero. Therefore, these connecting lines are not symmetric. We note that these complex solutions can not be obtained in a homogeneous medium and are therefore a manifestation of heterogeneity of the PC.

Referring to the data of Fig.~\ref{fig:IF_Circ}, we can conclude that, at the frequency $ka/\pi=0.200$, the dispersion relation in the PC is almost indistinguishable from the dispersion relation in a homogeneous effective medium with $n_{\rm eff}^2=6.81$. The correspondence holds well into the evanescent waves. Indeed, the range of $q_x$ shown in the figure for this frequency is $0 \leq q_x \leq 5 k$, and we can expect that the PC is homogenizable for $-5k \leq q_x \leq 5k$. In fact, the PC is homogenizable in an even wider interval of $q_x$. Indeed, the correspondence still holds in the ``reflected'' segments of the lines in the bottom plot. In the effective medium, these reflected segments correspond to $\vert q_x \vert > 5k$. The only slight discrepancy between the PC and the homogeneous medium can be observed near the edge of the FBZ ($q_x \approx 5k$). Overall, at this frequency, the dispersion relations in the PC mimic those of a homogeneous medium with a high precision and in a wide range of $q_x$.

Now let us move to Case (b), $ka/\pi = 0.680$. This frequency is slightly below the second $\Gamma$-point frequency. As one could expect from the theoretical arguments of Sec.~\ref{sec:gen}, the upper plot has a nearly circular, purely real lobe in the lower-left corner of the frame. If only this lobe is considered, one might erroneously conclude that, at $k a/\pi = 0.680$, the law of dispersion is almost the same as in an effective homogeneous medium. But a quick purview of the scale of the upper horizontal axis reveals that this isotropic behavior holds only in a very narrow range $-0.15 \lesssim q_x/k \lesssim 0.15$. In problems of transmission through a flat interface, this corresponds to incident angles of less than about $6^{\circ}$ with respect to the normal. Outside of the above range of $q_x$, the isotropy is lost as can be seen in the bottom plot. We emphasize that the isotropy is lost in this case for $\vert q_x \vert < k$. These values of $q_x$ correspond to propagating waves in a vacuum. It is true that these waves are evanescent in the PC but, upon transmission through a finite slab, incoming propagating waves are always transformed into outgoing propagating waves. Therefore, the transmission properties of the PC at this frequency are very different from those of any homogeneous medium. 

Note that, at sufficiently large values of $q_x$, the waves in PC switch from being evanescent to being propagating again (the upper-right corner lobe). This behavior is not characteristic of any homogeneous medium. Moreover, it can be seen that this PC is characterized by birefringence in some range of $q_x$ even though no anisotropy is involved - again, an effect not observed in homogeneous media. By birefringence we mean the effect when, for a given value of $q_x$, there exist Bloch waves with two real but different values of $q_y$, which corresponds to two different directions of propagation. This phenomenon can not occur in any local homogeneous medium.

In the Case (c), $ka/\pi=0.546$, there is obviously no hope to approximate the law of dispersion of the PC by the law of dispersion of any homogeneous effective medium. The real-valued lobe of the isofrequency line is severely distorted and birefringence is quite prominent and occurs in a wide range of $q_x$. As expected, the distortion of the real-valued lobe is consistent with $C_4$ symmetry of the problem. Note, however, that this type of distortion is suppressed in the triangular lattice, as will be shown below.

Next, we show the purely real quasi-circular lobes of the isofrequency lines in more detail in Fig.~\ref{fig:IF_Circ_Arcs}. It can be seen that the lobe is almost indistinguishable from a mathematical circle at the frequency $ka/\pi=0.200$. A distortion consistent with $C_4$ symmetry is clearly visible at the frequency $ka/\pi=0.680$. At $ka/\pi=0.546$ (far from the $\Gamma$-point), this distortion is quite severe.

Perhaps the most clear illustration of the departure from the homogeneous behavior in the cases (b) and (c) is shown in Fig.~\ref{fig:IF_Circ_Squares}, where we plot ${\rm Re}(q_y^2)$ as a function of $q_x^2$. In a homogeneous medium, this function is simply a straight line. This line is folded into the FBZ of the lattice as is shown in Fig.~\ref{fig:IF_Circ_Squares}(a) if the medium is artificially discretized (we note that for the particular parameters and the lattice type considered at the moment, the folding results only in linear segments; in a more general case the isofrequency line can acquire curvature due to the folding and an example will be shown below). This behavior is reproduced with very good accuracy at the frequency $ka/\pi=0.200$ [Case (a)]. However, in Cases (b) and (c) the departure from the linear behavior is obvious and dramatic. We will observe a similar behavior in the triangular lattice as well.

We finally note that the complex branches of the isofrequency lines that connect the purely imaginary and purely real segments of the data point sets is a peculiar feature of the PC, which can not be reproduced in any homogeneous medium with a real refractive index. In the latter case, $q_y$ is either purely real or purely imaginary and $q_y^2$ is always real.

\subsection{Square lattice with dispersive and absorbing host material}
\label{subsec:squre_disp}

To illustrate the effects of dispersion and absorption in the host material, we now assume that $\epsilon_h$  is a function of the frequency $\omega$ and is given by the following standard expression:
\begin{equation}
\label{eps_h_disp}
\epsilon_h(\omega) = 1 + [\epsilon_h(0) - 1] \frac{\omega_0^2}{\omega_0^2 - \omega^2 - i \gamma \omega} \ .
\end{equation}
\noindent
Here $\omega_0$ is the resonance frequency and $\gamma$ is the relaxation constant. In the simulations, we have taken $\epsilon_h(0) = 9.3628$, $\omega_0 a / \pi c = k_0h/\pi = 4$ and $\gamma/\omega_0 = 1/4\pi \approx 0.07958$. Then, at the frequency corresponding to $k a /\pi = 0.680$ (at which we compute the isofrequency line), we have $\epsilon_h \approx 9.61 + 0.12i$. The real part of this permittivity is the same as that of the non-absorbing host considered in the previous subsection.

\begin{figure}
\includegraphics[width=8.2cm]{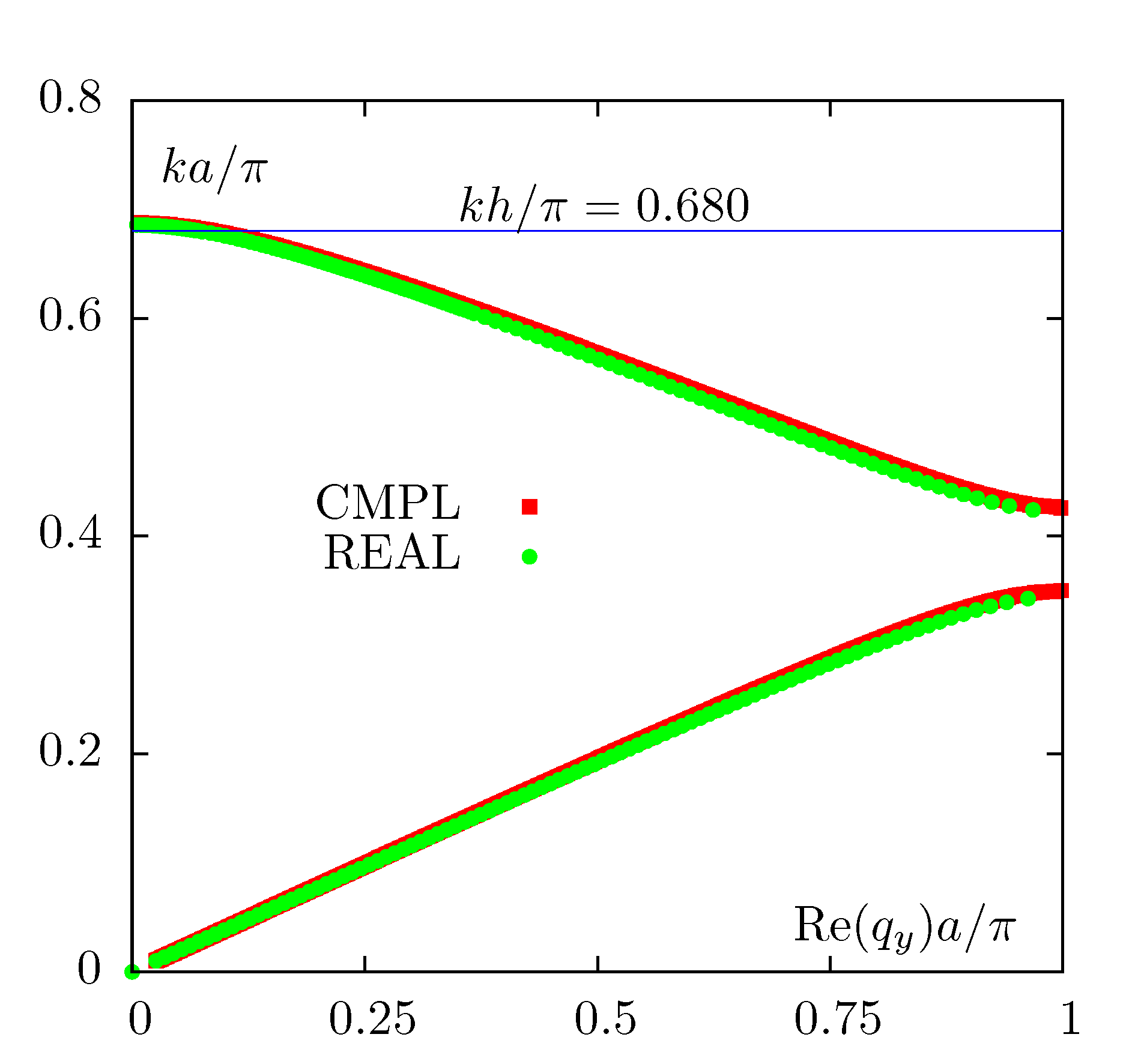}
\caption{(color online) \label{fig:Disp_CCir}
  Comparison of dispersion diagrams for the host material with a constant real permittivity $\epsilon_h=9.61$ (REAL) and with dispersive and complex permittivity given by \eqref{eps_h_disp} (CMPL). Geometrical parameters are the same as in Fig.~\ref{fig:Disp_Circ}. The horizontal line marks the frequency at which the isofrequency line has been computed for both cases. In the case of dispersive media, $q_y$ is complex and the quantity shown on the horizontal axis is ${\rm Re}(q_y)$.}
\end{figure}

The dispersion diagram of a PC with exactly the same geometry as above but with the host permittivity described by \eqref{eps_h_disp} is shown in Fig.~\ref{fig:Disp_CCir}. In this figure we compare the dispersion diagrams for the dispersive and non-dispersive hosts and it can be seen that they are barely distinguishable. The working frequency corresponding to $ka/\pi=0.680$ is still just below the $\Gamma$-point frequency in the second photonic band. However, the $\Gamma$-point frequency is not precisely defined for the dispersive medium because ${\bf q}$ is now complex at all frequencies. At the apparent $\Gamma$-point frequency that is visible in the figure, only ${\rm Re}({\bf q})$ turns to zero while ${\rm Im}({\bf q})$ does not. Correspondingly, the condition \eqref{Gamma_def_j>1} does not really hold at this frequency.

\begin{figure}
\includegraphics[width=8.2cm]{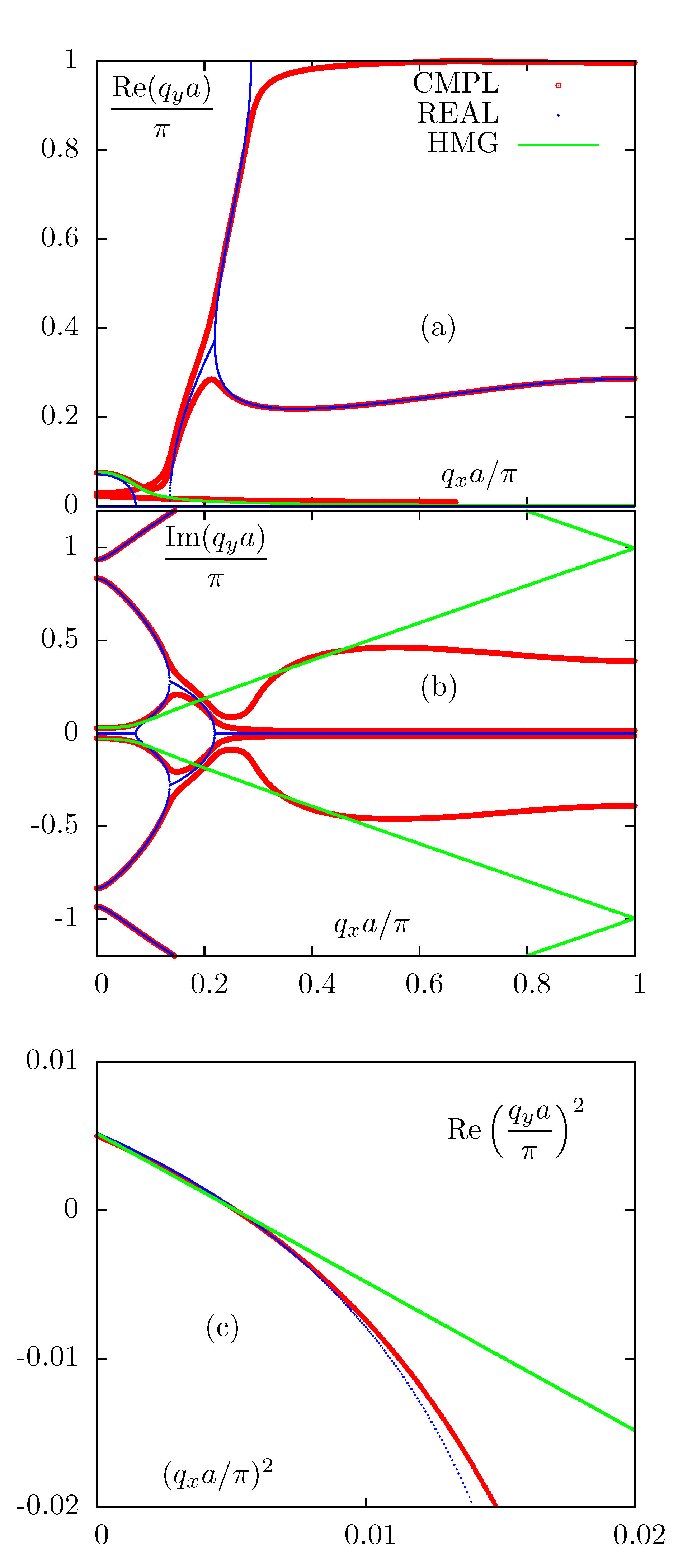}
\caption{(color online) \label{fig:IF_CCir}
 Isofrequency lines for the complex permittivity of the host $\epsilon_h = 9.61 + 0.12i$ and $ka/\pi=0.680$. Isofrequency lines of the absorbing PC are compared to the respective results for a PC with the purely real $\epsilon_h=9.61$ (thin blue line) and with the homogeneous effective medium with the refractive index $n_{\rm eff}^2=0.0117 - 0.00987i$ (intermediate green line). The top two panels are analogous to the column (b) in Fig.~\ref{fig:IF_Circ} (real and imaginary parts of $q_y$ as functions of $q_x$). The bottom panel is analogous to the column (b) in Fig.~\ref{fig:IF_Circ_Squares} (real part of $q_y^2$ as a function of $q_x^2$).}
\end{figure}

The isofrequency lines computed at $ka /\pi=0.680$ for the absorbing PC are shown in Fig.~\ref{fig:IF_CCir}. The lines are compared to those of a non-absorbing PC (thin blue line) and of a homogeneous effective medium whose refractive index $n_{\rm eff}$ was obtained by fitting the dispersion points to the analytical formula $q_y^2 = n_{\rm eff}^2 k^2 - q_x^2$ in the interval  $0\leq q_x a/\pi \leq 0.075$. The effective refractive index obtained from the above procedure is $n^2_{\rm eff} = 0.0117 - 0.00987i$. 

It can be seen from Fig.~\ref{fig:IF_CCir} that account of absorption does not have a significant effect on the homogenizability of the PC. The analysis, however, becomes more complicated because all values of $q_y$ are now complex. An interesting result that is not directly relevant to homogenization is the following. In the case of non-absorbing host, there exists a region of $q_x$ for which $q_y$ has nonzero real and imaginary parts [see Fig.~\ref{fig:IF_Circ}(b)]. The corresponding segment of the isofrequency line connects the purely imaginary solutions to the purely real lobe in the upper-right corner of the plot. In the case of an absorbing host, these complex solutions split into two distinct branches with slightly different imaginary and real parts; the upper-right lobe is also deformed and split into two disjoint segments. As a result, absorption in the host material enhances the effect of birefringence. for almost all values of $q_x$, the PC will transmit two distinct Bloch waves whose rate of spatial decay is not very different. Of course, there will also be an infinite number of Bloch waves with very fast spatial decay.

The above observation that non-zero absorption results in splitting of the isofrequency line into distinct and disconnected branches may be interesting and deserving an additional investigation, but it has no direct bearing on the main subject of this paper, which is homogenization. At least, the appearance of birefringence makes homogenization only harder.

\subsection{Triangular lattice with non-dispersive and non-absorbing host}
\label{subsec:tri}

We now turn to the case of the triangular lattice. The primitive vectors for the real-space and reciprocal lattices are 
\begin{eqnarray*}
&& {\bf a}_1 = a(1,0) \ , \  {\bf a}_2 = a(1/2,\sqrt{3}/2) \ , \\
&& {\bf b}_1 = \dfrac{2\pi}{a}(1,-1/\sqrt{3}) \ , \ \ {\bf b}_2 = \dfrac{2\pi}{a}(0,2/\sqrt{3}) \ .
\end{eqnarray*}
\noindent
At the center of each elementary cell, which is now rhombic, we place a hollow cylinder of the radius $R=0.42a$. These inclusions do not cross the boundaries of the elementary cell. It can be seen that the inclusions form a triangular lattice. Since the inclusion obeys the same symmetry as the lattice, the whole structure is $C_6$-symmetric. The permittivity of the host matrix is taken to be $\epsilon_h=12.25$. This model was used in Ref.~\onlinecite{pei_12_1}. 

The FBZ of the lattice described above is a rhombus and any of the two triangles created by drawing a rhombus diagonal are equivalent. One can construct a hexagon of two complete rhombuses touching at one corner and two such triangles. All six triangles forming this hexagon are equivalent and, moreover, the dispersion points can be replicated periodically on the hexagonal lattice. It is a common practice to plot the isofrequency lines of such lattices inside a hexagonal region of the reciprocal lattice. However, we do not use this approach in the paper and limit attention to the actual FBZ of the lattice, which is shown in all figures below.

\begin{figure}
\includegraphics[width=8.2cm]{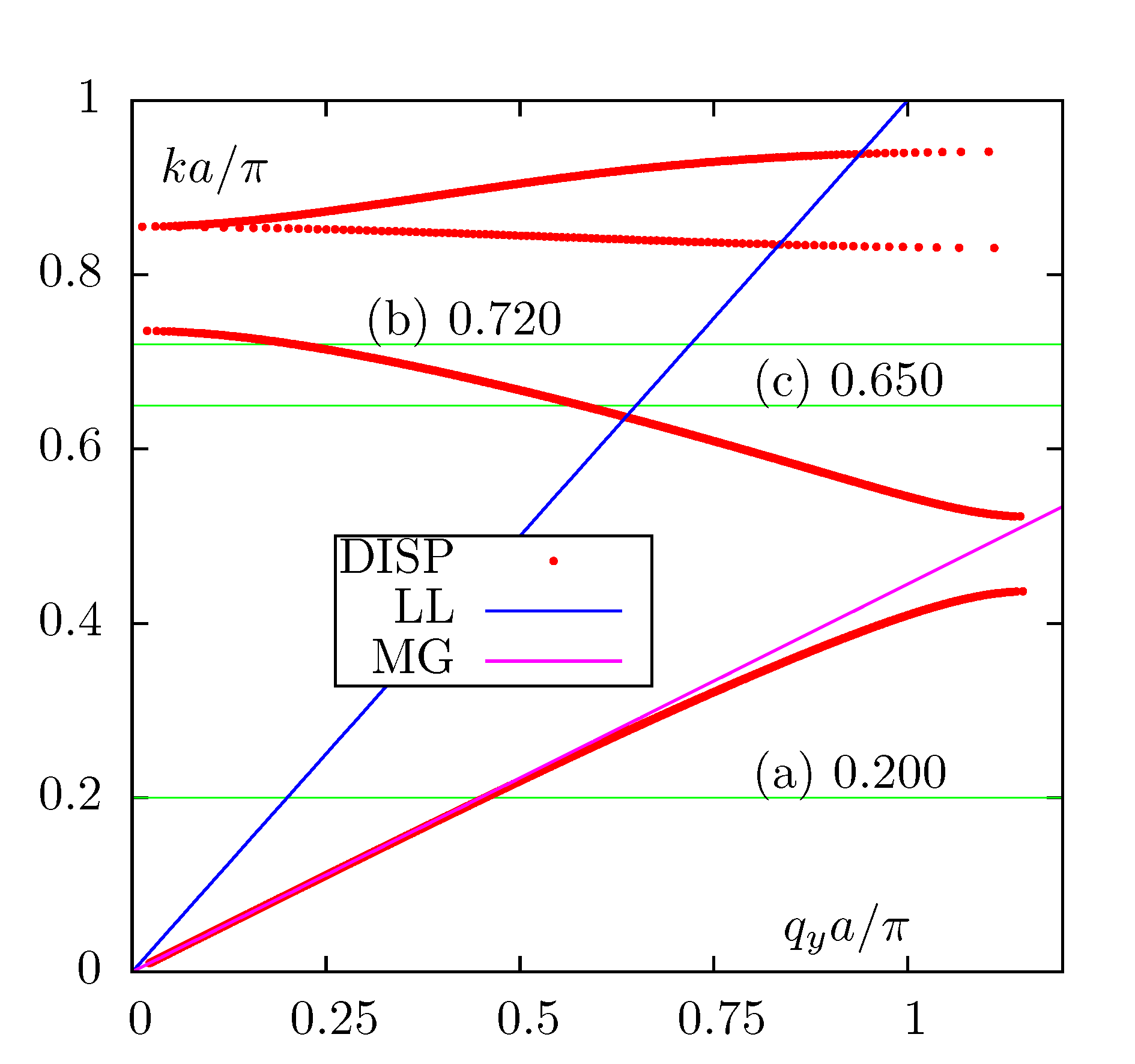}
\caption{(color online) \label{fig:Disp_TCir}
  Same as in Fig.~\ref{fig:Disp_Circ} but for the triangular lattice of hollow cylinders of the radius $R=0.42a$ in the host medium of the permittivity $\epsilon=12.25$.}
\end{figure}

The dispersion diagram for the direction of propagation along the $Y$-axis ($q_x=0$) is shown in Fig.~\ref{fig:Disp_TCir}. As previously, three special frequencies are marked by the horizontal lines in Fig.~\ref{fig:Disp_TCir}. These frequencies are $ka/\pi=0.200$ [Case (a)], $ka/\pi=0.720$ [Case (b)] and $ka/\pi=0.650$ [Case (c)]. Just as was the case for the square lattice, the frequency (a) is in the first photonic band, the frequency (b) is in the second band slightly below the second $\Gamma$-point frequency and the frequency (c) is close to the intersection of the light line and the second branch of the dispersion curve.

\begin{figure*}
\includegraphics[width=16.4cm]{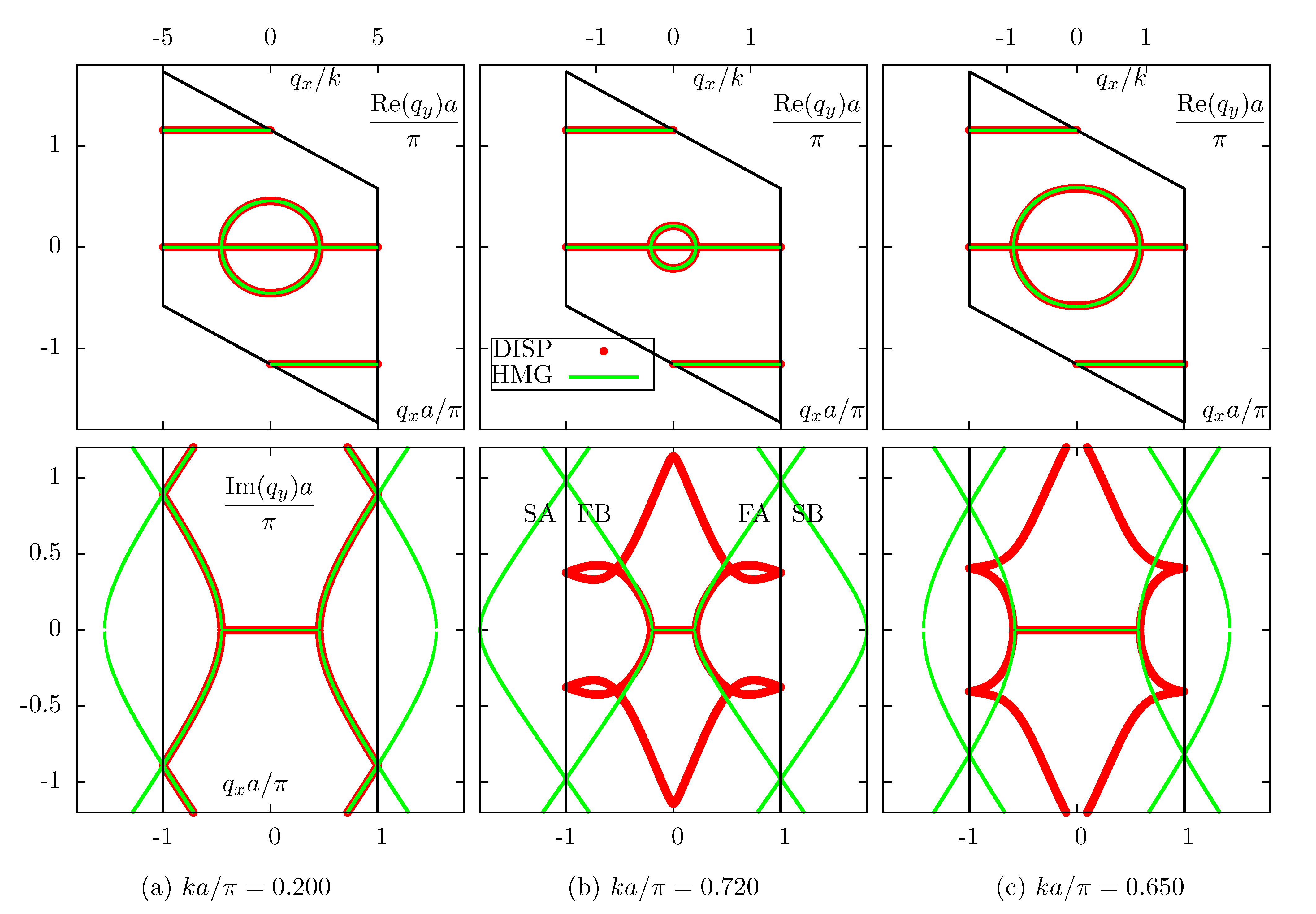}
\caption{(color online) \label{fig:IF_TCir} Same as in Fig.~\ref{fig:IF_Circ} but for the triangular lattice of hollow cylinders. The 'HMG' isofrequency lines were computed according to \eqref{IFE_HMG_PER} with $n_{\rm eff}^2 = 5.20$ (a),  $n_{\rm eff}^2 = 0.085$ (b) and $n_{\rm eff}^2 = 0.82$ (c). In the top row of plots, the FBZ of the lattice is shown by the black rhombus. In the lower plot (b), FA and FB mark the first pair of hyperbolas defined by \eqref{tri_fold_im_qy_qx}, and SA, SB mark the second pair of hyperbolas.}
\end{figure*}

The complex isofrequency lines for the three frequencies mentioned above are shown in Fig.~\ref{fig:IF_TCir}. This figure is analogous to Fig.~\ref{fig:IF_Circ} and the same quantities are plotted using the same scales of the axes. However, the FBZ of a triangular lattice is more complex geometrically. Therefore, in the top row of plots of Fig.~\ref{fig:IF_TCir}, we have shown the complete FBZ of the lattice (the black rhombus) while in Fig.~\ref{fig:IF_Circ}, only one quarter of the FBZ was shown.

We now analyze Fig.~\ref{fig:IF_TCir} in more detail. First, focus on the real parts of $q_y$ (the upper row of plots). The central quasi-circular lobe and the horizontal line ${\rm Re}(q_y) = 0$ are analogous to the similar features of the isofrequency lines for a square lattice. The quasi-circular lobes consist of purely real solutions while the line $q_y=0$ corresponds to purely imaginary solutions. The upper and lower horizontal lines in the upper row of plots correspond to complex solutions that are specific to the triangular lattice. Unlike in the case of a square lattice, these complex solutions do not connect purely imaginary and purely real segments of the isofrequency line (within the FBZ). Another distinction is that the complex solutions in a triangular lattice will appear due to folding of the dispersion equation of an artificially discretized homogeneous medium. For a square lattice, the appearance of complex solutions is a result of interaction that can not be obtained by artificial folding.

We can understand the appearance of the complex solutions shown in Fig.~\ref{fig:IF_TCir} by replicating the central rhombus of the top row of plots in all directions and noting that the horizontal line $q_y=0$ in the central rhombus will connect to the upper or lower horizontal lines in the replicated rhombuses. We can also understand these solutions qualitatively by considering a homogeneous medium artificially discretized on a triangular lattice. In the infinite (non-discretized) medium, the law of dispersion allows for purely imaginary solutions of the form $q_y = \pm i \sqrt{q_x^2 - n^2 k^2}$ for $\vert q_x \vert > nk$, $n>0$ being the index of refraction. Now let us fold these solutions to the FBZ of a triangular lattice according to \eqref{IFE_HMG_PER}. We can ignore the imaginary part of ${\bf q}$ for the purpose of computing the integers $n_1$, $n_2$ that are used in this equation. The result of the folding is
\begin{subequations}
\label{tri_fold}
\begin{align}
\label{tri_fold_qx}
&\left[ q_x \right]_{\rm FBZ} = q_x - \frac{2\pi}{a} n_1  \ , \\
\label{tri_fold_qy}
&\left[ q_y \right]_{\rm FBZ} = \pm i \sqrt{q_x^2 - n^2 k^2} - \frac{2\pi}{a\sqrt{3}} \left(2 n_2 - n_1 \right) \ , \\
& n_1 = {\rm Nint}\left( \frac{q_x a}{2\pi} \right) \ , \ \ n_2 = {\rm Nint}\left( \frac{q_x a}{4\pi} \right) \ .
\end{align}
\end{subequations} 
\noindent
These equations are valid for $\vert q_x \vert > nk$. We note that the integer index $m = 2n_2 - n_1$ can take only three values: $0,\pm 1$. Correspondingly, for the real and imaginary parts of $\left[ q_y \right]_{\rm FBZ}$, we have the following results:
\begin{subequations}
\label{tri_fold_re_im_qy}
\begin{align}
\label{tri_fold_re_qy}
& \left[ {\rm Re}(q_y) \right]_{\rm FBZ} = m \frac{2\pi}{a\sqrt{3}} \ , \ \ m = 0,\pm 1 \ , \\
\label{tri_fold_im_qy}
& \left[ {\rm Im}(q_y) \right]_{\rm FBZ} =  \pm \sqrt{q_x^2 - n^2 k^2} \ .
\end{align}
\end{subequations} 
\noindent
We will obtain a dispersion equation containing only the quantities $\left[ q_x \right]_{\rm FBZ}$ and $\left[ q_y \right]_{\rm FBZ}$ if we substitute $q_x$ in \eqref{tri_fold_im_qy} from \eqref{tri_fold_qx}, i.e., use the relation
\begin{equation}
q_x^2 =  \left\{ \left[ q_x \right]_{\rm FBZ} + \frac{2\pi}{a} n_1 \right\}^2 
\end{equation}
\noindent
to obtain the following closed-form equation:
\begin{eqnarray}
\label{tri_fold_im_qy_qx}
\left[ {\rm Im}(q_y) \right]_{\rm FBZ} =  \pm \sqrt{\left\{ \left[ q_x \right]_{\rm FBZ} + \frac{2\pi}{a} n_1 \right\}^2 - n^2 k^2} \\
\nonumber
{\rm for} \ \ \left\vert \left[ q_x \right]_{\rm FBZ} + \frac{2\pi}{a} n_1 \right\vert \geq n k \ .
\end{eqnarray} 
\noindent
This equation describes a family of hyperbolas parameterized by $n_1$. The first pair of these hyperbolas (corresponding to $n_1=0$) are labeled as FA and FB in Fig.~\ref{fig:IF_TCir}. The second pair or hyperbolas (including two branches with $n_1=\pm1$) are labeled as SA and SB. Infinitely many similar curves can be generated by translations along the $q_x$-axis, which correspond to arbitrary integer values of $n_1$. Note that the two integers $n_1$ and $n_2$ that label different ``reflected'' branches of the solution to the dispersion equation are subject to the selection rule $m = n_1 - 2 n_2 = 0,\pm 1$. Therefore, for each $n_1$, the set of allowable values of $n_2$ is restricted.

Thus, the FBZ folding in a triangular lattice transforms purely imaginary solutions $q_y$ into complex solutions $\left[ q_y \right]_{\rm FBZ}$, which explains the appearance of the upper and lower horizontal lines in the top row of plots in Fig.~\ref{fig:IF_TCir}. 

\begin{figure}
\includegraphics[width=8.2cm]{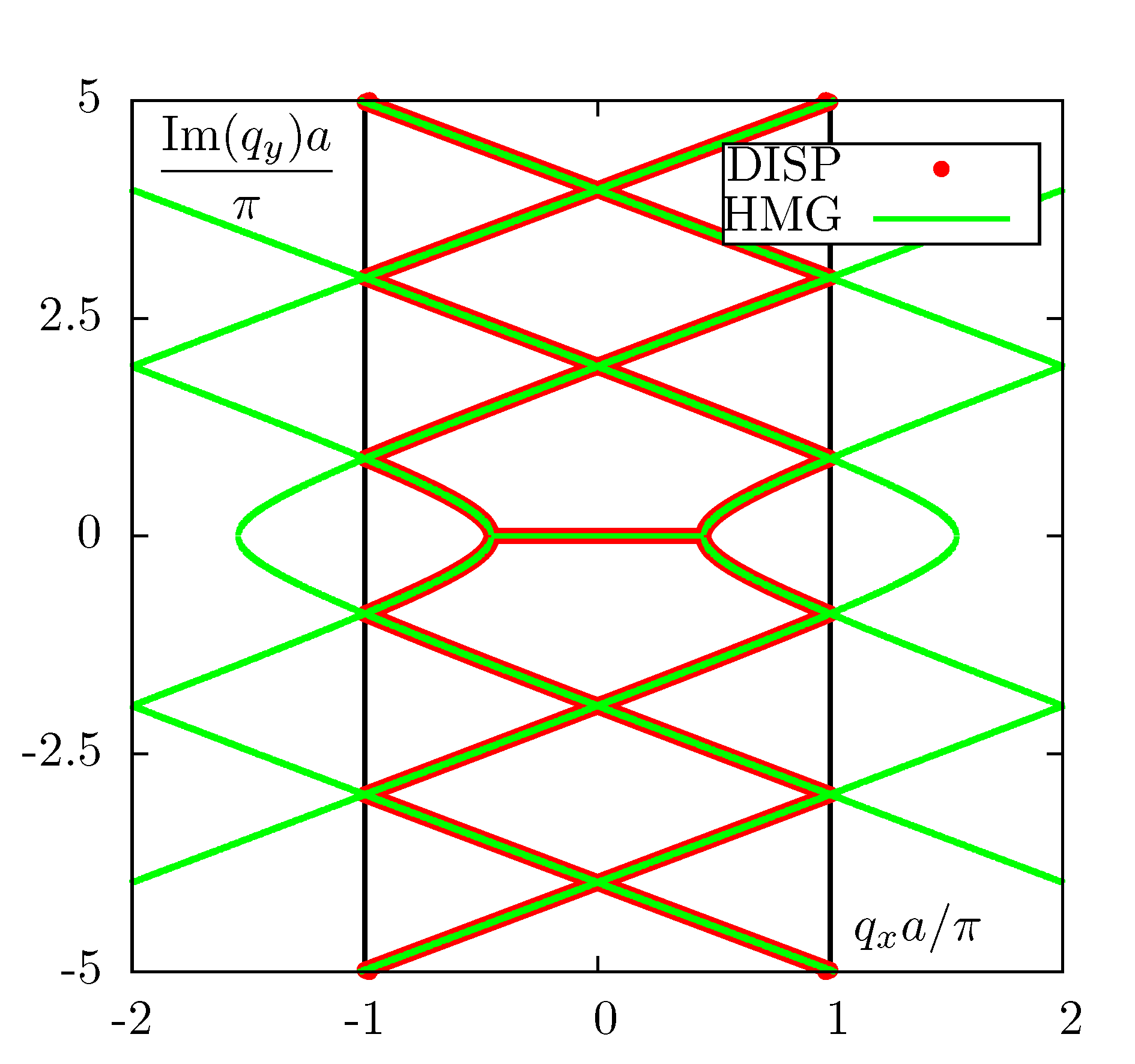}
\caption{(color online) \label{fig:IF_TCir_HMG}
  An expanded view of ${\rm Im}(q_y)$ as a function of $q_x$ for $ka/\pi=0.200$ [Case (a)]. Here several 'reflected' segments of the isofrequency line are shown. These segments are obtained by folding the corresponding isofrequency line for an infinite homogeneous medium into the FBZ of the triangular lattice according to \eqref{tri_fold_im_qy_qx}. The analytical lines outside the FBZ (shown by the two vertical lines) are not the actual solutions and are shown only to guide the eye.}
\end{figure}

Of course, the analytical folding described above is valid only for a homogeneous medium. However, at the frequency $ka/\pi=0.200$, the actual dispersion relation in the PC mimics the dispersion relation in a homogeneous effective medium with $n_{\rm eff}^2=5.20$ very closely. This conclusion can be drawn from the data shown in the lower plot for the Case (a) in Fig.~\ref{fig:IF_TCir} and is further confirmed and illustrated in Fig.~\ref{fig:IF_TCir_HMG}. In the latter figure, we plot $\left[{\rm Im}(q_y)\right]_{\rm FBZ}$ as a function of $\left[ q_x \right]_{\rm FBZ}$ (the symbol $[ \ldots ]_{\rm FBZ}$ that signifies Bloch-periodicity of all data points is omitted in this and other figures for brevity). The comparison is made between the respective quantities obtained for the actual PC and a homogeneous effective medium artificially discretized on the same lattice. The individual hyperbolas described by \eqref{tri_fold_im_qy_qx} are shown with green lines and the actual dispersion solutions in the PC are shown by the red points. Both solutions are only valid within the FBZ (between the two vertical lines) and beyond these two lines they must be periodically replicated. The green lines outside of the FBZ are shown in the figure only to guide the eye (to help visually identify individual hyperbolas).

It can be seen that, at the dimensionless frequency $ka/\pi=0.200$, the law of dispersion in the PC is almost indistinguishable from the law of dispersion in a homogeneous medium, at least up to $\left \vert {\rm Im}(q_y) a / \pi \right \vert \leq 5$, which corresponds, approximately, to $\left\vert q_x \right\vert \lesssim 25k$, that is, very far into the evanescent spectrum. Therefore, we can claim that, at this particular frequency, the PC can be homogenized for many practical purposes.

The situation is quite different at the other frequencies considered. In Case (b) ($ka/\pi=0.720$, just below the second $\Gamma$-point frequency), the dependence of ${\rm Re}(q_y)$ on $q_x$ still looks very ``homogeneous''. However, as soon as we look at ${\rm Im}(q_y)$, it becomes obvious that the law of dispersion departs from that of a homogeneous medium quite dramatically as soon as $q_x$ approaches the region of evanescent waves ($\vert q_x \vert \gtrsim k$). Definitely, homogenization is not possible for incident evanescent waves, and it is inaccurate for propagating waves with large angles of incidence, e.g., for $\theta_{\rm inc} \gtrsim 70^\circ$. For the frequency (c) [$ka/\pi = 0.650$], the quasi-circular lobe is much larger and appears to not be distorted significantly. However, the lower plot again clearly indicates a lack of correspondence between the law of dispersion of a homogeneous medium and the law of dispersion in the PC.

We therefore conclude that the PC is not homogenizable at the frequencies (b) and (c). Assigning the PC some effective parameters at these frequencies can potentially be a valid approximation only for a limited range of incident angles that, at the very least, do not include evanescent waves. 

\begin{figure*}
\includegraphics[width=16.4cm]{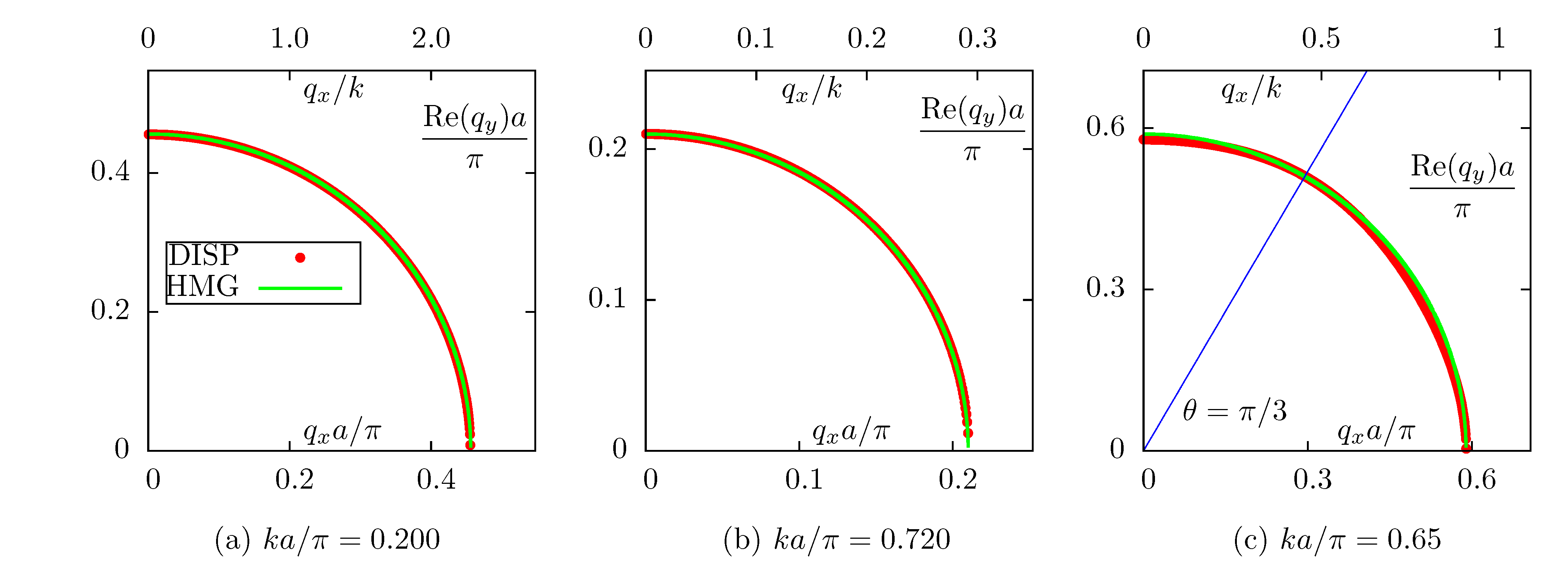}
\caption{(color online) \label{fig:IF_TCir_Arcs} Same as in Fig.~\ref{fig:IF_Circ_Arcs} but for the triangular lattice. Distortions of the quasi-circular lobes are consistent with $C_6$ symmetry. The straight line in panel (c) makes the $\pi/3$ angle with the horizontal axis.}
\end{figure*}

Next, as was done above for the square lattice, we show in Fig.~\ref{fig:IF_TCir_Arcs} the quasi-circular lobes of the top row of Fig.~\ref{fig:IF_TCir} in more detail. In order to make the small deviations from circularity more visible, we plot in this figure only one quarter of each quasi-circular lobe. The lobes appear to be indistinguishable from mathematical circles in Cases (a) and (b) but some small distortions consistent with the $C_6$ symmetry are visible in Case (c). The high quality of the quasi-circular lobes can be explained by the fact that the terms of the form \eqref{C4} do not enter the expansion of $\Sigma(\omega,{\bf q})$ because they are not invariants of $C_6$ symmetry. Therefore, the distortions first appear in six-order terms \eqref{C6}. However, as soon as the terms \eqref{C6} become non-negligible, the isotropy of $\Sigma(\omega, {\bf q})$ is lost quickly. Therefore, the high quality of the quasi-circular lobes shown in Fig.~\ref{fig:IF_TCir_Arcs} is not a sufficient condition for homogenizability.

\begin{figure*}
\includegraphics[width=16.4cm]{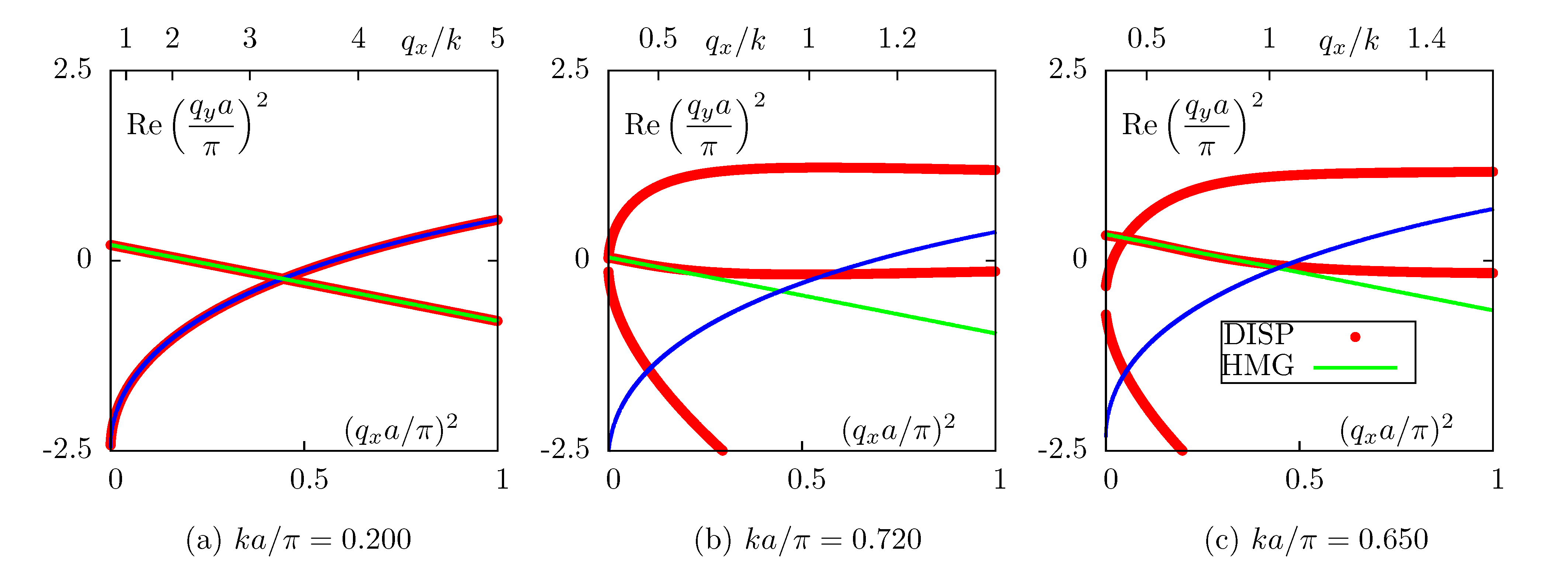}
\caption{(color online) \label{fig:IF_TCir_Squares} Same as in Fig.~\ref{fig:IF_Circ_Squares} but for the triangular lattice that is considered in this subsection.}
\end{figure*}

But the most convincing demonstration of non-homogenizability of the PC at the frequencies (b) and (c) can be obtained by considering the squares of the Cartesian components of ${\bf q}$. In Fig.~\ref{fig:IF_TCir_Squares}, we plot ${\rm Re}(q_y^2)$ vs $q_x^2$. In a homogeneous effective medium artificially discretized on a triangular lattice, this plot consists of one linear and several curved segments. The linear segment corresponds to the eigenvalues that are either purely real or purely imaginary. The curved segments correspond to the peculiar complex eigenvalues that are specific to the triangular lattice and described mathematically by the equations \eqref{tri_fold} through \eqref{tri_fold_im_qy_qx}. One such curved segment is shown in every panel of Fig.~\ref{fig:IF_TCir_Squares}; other curved segments lie outside of the plot frames. We now observe that at the frequency (a) there is a complete correspondence between the artificially discretized homogeneous medium and the PC. This, of course, could have been expected from the data shown in Figs.~\ref{fig:IF_TCir}, \ref{fig:IF_TCir_HMG},\ref{fig:IF_TCir_Arcs}. At the frequencies (b) and (c), the correspondence is broken in the linear segment. In addition the curved segments in the homogeneous medium and in the PC are completely different. The last point is significant and deserves an additional discussion.

Let us assume that a slab of a {\em homogeneous} material contained between two planes $y=0$ and $y=L$ is illuminated by a plane wave with the incidence angle such that $q_x < \pi/a$, where $a$ is the period of artificial discretization in the $x$-direction. We can describe the medium as homogeneous (the traditional approach) or as a PC (by using artificial discretization). Both approaches are mathematically equivalent and will yield the same results for all observables. Assume that we have decided to describe the medium as a PC. In this case, for a given purely real $q_x$, there will be infinitely many eigenvalues $q_y$. However, the incident radiation will excite only one mode, namely, the mode with $q_y$ that corresponds to the dispersion relation of the homogeneous material. The modes with other values of $q_y$ can be excited if we take $q_x$ to be outside of the FBZ of the lattice. In any case, for a given $q_x$, only one mode is excited in the material.

In the case of an actual PC whose law of dispersion closely mimics that of a homogeneous medium, as was the case at the frequency (a), we can expect that the same selection rules will work: at any given $q_x$, only one mode will be excited in the PC. Then the transmission and reflection coefficients can be expected to be the same in the PC and the homogeneous medium. This is indeed the case in the homogenization limit~\cite{markel_12_1,tsukerman_14_1} $a\rightarrow 0$. 

If we consider the PC with a constant nonzero $a$ at sufficiently high excitation frequencies (such as the frequencies (b) and (c) in the above examples), there is no reason to believe that the selection rules will work the same way as in a homogeneous medium. In other words, at a given $q_x$, modes with several different values of $q_y$ can be excited in the PC. Granted, these additional modes have $q_y$ with nonzero imaginary parts and are therefore exponentially decaying inside the medium. However, it is a mistake to neglect these modes completely. Indeed, even the fields associated with these modes are exponentially decaying with $y$, they are not negligibly small at the interface $y=0$. Moreover, these modes do not generally average to zero over the surface of the elementary cell. In these respect, they are different from the surface waves discussed by us previously~\cite{markel_12_1,xiong_13_1}. Therefore, excitation of these additional modes will have an adverse effect on homogenizability.

In Fig.~\ref{fig:IF_TCir_Squares}, two different effects are illustrated. The first effect is the deviation of the law of dispersion in the PC from that in a homogeneous effective medium for the ``fundamental mode'' (for which $q_y^2$ is always purely real). This effect is manifest at sufficiently large incident angles and, in particular, for incident evanescent waves. The second, more subtle effect is manifest even at small incident angles. Namely, it can be seen that, at $q_x \sim 0$, the PC has additional modes (allowable values of $q_y$) that are dramatically different from the respective values in the effective medium. Appearance of these additional solutions can be expected to influence the medium impedance in an angle-dependent manner and have an additional degrading effect on the PC homogenizability.

\section{Discussion}
\label{sec:disc}

The main message of this paper is that, in many applications of practical interest, it is insufficient to consider only the purely real segments of the isofrequency lines to decide whether a given photonic crystal (PC) is homogenizable - that is, electromagnetically similar to a homogeneous medium. In the case of two-dimensional PCs that we have considered, these purely real segments can be circular with reasonable precision in the higher photonic bands. In this case, all propagation directions appear to be equivalent and the physical effects of the medium discreteness (the presence of the lattice) appear to be minimized or absent. The medium can also be characterized by negative dispersion, which means that the real part of the Bloch wave number tends to decrease with frequency while the imaginary part is negligibly small. Nevertheless, the PC is not homogenizable in this case and can not be characterized by angle-independent, purely local effective parameters $\epsilon_{\rm eff}$ and $\mu_{\rm eff}$. This conclusion can be drawn by considering the angular dependence of the effective parameters and by including evanescent waves into consideration. 

A difficulty one faces when restricting consideration to the dispersion relation is that infinite media have no interfaces and therefore it is easy to overlook the important features of the solutions. This is exactly what happens when one restricts consideration to purely real isofrequency lines. In this paper, we have generalized this approach by considering two orthogonal directions in space, $X$ and $Y$, and assuming that the projection of the Bloch wave vector ${\bf q}$ on one of these axes ($X$ in our case) is a mathematically-independent variable, which is preserved by the process of reflection and refraction at any planar interface $y = {\rm const}$.

The results obtained above are consistent with an earlier prediction made by one of the authors regarding the impossibility of negative refraction~\cite{markel_08_1}. The essential assumption of the above reference was that the medium in question is electromagnetically homogeneous. However, this requirement was not clearly defined. On the other hand, it is a common knowledge that some discrete systems such as photonic crystals or chains of interacting particles or resonators can be characterized by {\em negative dispersion}. This is true, in particular, for PCs in higher photonic bands. This may seem to contradict the conclusions of Ref.~\onlinecite{markel_08_1}. In the present work, we show that there is no contradiction since PCs are not homogenizable in the higher bands. We also define more clearly what we mean by the requirement that the medium is ``electromagnetically homogeneous''. Specifically, we have formulated a {\em necessary} condition of homogenizability of a heterogeneous medium. This condition is based on the correspondence of the dispersion relations in a hypothetical homogeneous effective medium and the actual PC. A {\em sufficient} condition must include the impedance~\cite{tsukerman_14_1}, which can not be defined unambiguously without explicit consideration of the medium boundary~\cite{markel_13_1}. Thus, we do not discuss a sufficient condition of homogenizability in this paper. Instead, we show that even the necessary condition is violated in the higher photonic bands. 

Note that the lack of homogenizability at sufficiently high frequencies was demonstrated earlier for the special case of 1D periodic media~\cite{markel_10_1}. The latter case is, however, complicated by the fact that a 1D layered medium is always anisotropic. Certain types of indefinite anisotropic media are capable of refracting a narrow incident beam on the same side of the normal. This phenomenon can easily be confused with negative refraction. In this paper, we consider a system without anisotropy (either magnetic or electric). 

A more general result closely related to the main conclusion of this paper can be stated in the form of an {\em uncertainty principle} of homogenization~\cite{tsukerman_16_1}, namely: the more an effective magnetic permeability deviates from unity, the less accurate the corresponding homogenization result is. In the former reference as well as in this paper, we define the effective medium parameters $\epsilon_{\rm eff}$ and $\mu_{\rm eff}$ by requiring that a heterogeneous and an ``effective'' homogeneous samples (of the same overall shape) can not be distinguished by performing {\em external} measurements of reflected, transmitted and scattered waves in a sufficiently broad range of illumination conditions. We have previously formulated this requirement quantitatively for the case of a plane-parallel slab~\cite{tsukerman_14_1}. We believe, however, that, if the medium is truly homogenizable, the effective parameters can also be used to compute various physical quantities that are {\em internal} to the medium such as dissipated heat per unit volume per unit time. These quantities are usually quadratic in the fields. In particular, we believe that the effective permittivity $\epsilon_{\rm eff}$, when it can be introduced, satisfies the usual physical requirements of passivity and causality. According to the homogenization uncertainty principle~\cite{tsukerman_16_1}, the effective permeability $\mu_{\rm eff}$ can not be much different from unity, and the physical requirements applicable to this quantity may be more nuanced~\cite{markel_08_2}. We note that these questions are of current interest in the theory of homogenization but, unfortunately, outside of the scope of this paper wherein we do not compute or consider spatial distribution and fluctuations of the electromagnetic fields in a PC, or the medium boundaries.

In Appendix~\ref{app:A}, we explain the mathematical distinction between spurious $\Gamma$-point frequencies that are due to Brillouin-zone folding of Bloch bands and ``genuine'' $\Gamma$-point frequencies that are due to multiple scattering. Understanding this distinction is important for the theory of homogenization. 

We finally note that this work is based upon some important theoretical observations made by Li, Holt and Efros~\cite{li_06_1}. We have developed these observations further with a specific focus on dispersion relations in 2D PCs with square and triangular lattices.

This work has been carried out thanks to the support of the A*MIDEX project (No. ANR-11-IDEX-0001-02) funded by the ``Investissements d'Avenir'' French Government program, managed by the French National Research Agency (ANR) and was also supported by the US National Science Foundation under Grants DMS1216970 and DMS1216927.

\bibliographystyle{apsrev}
\bibliography{abbrev,master,book,local}

\begin{thebibliography}{41}
\expandafter\ifx\csname natexlab\endcsname\relax\def\natexlab#1{#1}\fi
\expandafter\ifx\csname bibnamefont\endcsname\relax
  \def\bibnamefont#1{#1}\fi
\expandafter\ifx\csname bibfnamefont\endcsname\relax
  \def\bibfnamefont#1{#1}\fi
\expandafter\ifx\csname citenamefont\endcsname\relax
  \def\citenamefont#1{#1}\fi
\expandafter\ifx\csname url\endcsname\relax
  \def\url#1{\texttt{#1}}\fi
\expandafter\ifx\csname urlprefix\endcsname\relax\def\urlprefix{URL }\fi
\providecommand{\bibinfo}[2]{#2}
\providecommand{\eprint}[2][]{\url{#2}}

\bibitem[{\citenamefont{Hasar et~al.}(2014)\citenamefont{Hasar, Buldu, Bute,
  Barroso, Karacali, and Ertugrul}}]{hasar_14_1}
\bibinfo{author}{\bibfnamefont{U.~C.} \bibnamefont{Hasar}},
  \bibinfo{author}{\bibfnamefont{G.}~\bibnamefont{Buldu}},
  \bibinfo{author}{\bibfnamefont{M.}~\bibnamefont{Bute}},
  \bibinfo{author}{\bibfnamefont{J.~J.} \bibnamefont{Barroso}},
  \bibinfo{author}{\bibfnamefont{T.}~\bibnamefont{Karacali}}, \bibnamefont{and}
  \bibinfo{author}{\bibfnamefont{M.}~\bibnamefont{Ertugrul}},
  \bibinfo{journal}{AIP Advances} \textbf{\bibinfo{volume}{4}},
  \bibinfo{pages}{107116} (\bibinfo{year}{2014}).

\bibitem[{\citenamefont{Karamanos et~al.}(2014)\citenamefont{Karamanos,
  Assimonis, Dimitriadis, and Kantartzis}}]{karamanos_14_1}
\bibinfo{author}{\bibfnamefont{T.~D.} \bibnamefont{Karamanos}},
  \bibinfo{author}{\bibfnamefont{S.~D.} \bibnamefont{Assimonis}},
  \bibinfo{author}{\bibfnamefont{A.~I.} \bibnamefont{Dimitriadis}},
  \bibnamefont{and} \bibinfo{author}{\bibfnamefont{N.~V.}
  \bibnamefont{Kantartzis}}, \bibinfo{journal}{Photonics and Nanostructures}
  \textbf{\bibinfo{volume}{12}}, \bibinfo{pages}{291} (\bibinfo{year}{2014}).

\bibitem[{\citenamefont{Clausen et~al.}(2014)\citenamefont{Clausen, Arslanagic,
  and Breinbjerg}}]{clausen_14_1}
\bibinfo{author}{\bibfnamefont{N.~C.~J.} \bibnamefont{Clausen}},
  \bibinfo{author}{\bibfnamefont{S.}~\bibnamefont{Arslanagic}},
  \bibnamefont{and}
  \bibinfo{author}{\bibfnamefont{O.}~\bibnamefont{Breinbjerg}},
  \bibinfo{journal}{Photonics and Nanostructures}
  \textbf{\bibinfo{volume}{12}}, \bibinfo{pages}{419} (\bibinfo{year}{2014}).

\bibitem[{\citenamefont{Ciattoni and Rizza}(2015)}]{ciattoni_15_1}
\bibinfo{author}{\bibfnamefont{A.}~\bibnamefont{Ciattoni}} \bibnamefont{and}
  \bibinfo{author}{\bibfnamefont{C.}~\bibnamefont{Rizza}},
  \bibinfo{journal}{Phys. Rev. B} \textbf{\bibinfo{volume}{91}},
  \bibinfo{pages}{184207} (\bibinfo{year}{2015}).

\bibitem[{\citenamefont{Sozio et~al.}(2015)\citenamefont{Sozio, Vallecchi,
  Albani, and Capolino}}]{sozio_15_1}
\bibinfo{author}{\bibfnamefont{V.}~\bibnamefont{Sozio}},
  \bibinfo{author}{\bibfnamefont{A.}~\bibnamefont{Vallecchi}},
  \bibinfo{author}{\bibfnamefont{M.}~\bibnamefont{Albani}}, \bibnamefont{and}
  \bibinfo{author}{\bibfnamefont{F.}~\bibnamefont{Capolino}},
  \bibinfo{journal}{Phys. Rev. B} \textbf{\bibinfo{volume}{91}},
  \bibinfo{pages}{205127} (\bibinfo{year}{2015}).

\bibitem[{\citenamefont{Pendry}(2000)}]{pendry_00_1}
\bibinfo{author}{\bibfnamefont{J.~B.} \bibnamefont{Pendry}},
  \bibinfo{journal}{Phys. Rev. Lett.} \textbf{\bibinfo{volume}{85}},
  \bibinfo{pages}{3966} (\bibinfo{year}{2000}).

\bibitem[{\citenamefont{Pokrovsky and Efros}(2002)}]{pokrovsky_02_1}
\bibinfo{author}{\bibfnamefont{A.~L.} \bibnamefont{Pokrovsky}}
  \bibnamefont{and} \bibinfo{author}{\bibfnamefont{A.~L.} \bibnamefont{Efros}},
  \bibinfo{journal}{Solid State Comm.} \textbf{\bibinfo{volume}{124}},
  \bibinfo{pages}{283} (\bibinfo{year}{2002}).

\bibitem[{\citenamefont{Rautian}(2008)}]{rautian_08_1}
\bibinfo{author}{\bibfnamefont{S.~G.} \bibnamefont{Rautian}},
  \bibinfo{journal}{Phys. Usp.} \textbf{\bibinfo{volume}{51}},
  \bibinfo{pages}{981} (\bibinfo{year}{2008}).

\bibitem[{fn1()}]{fn1}
\bibinfo{note}{There can be no electric anisotropy because the electric field
  in the considered geometry is always aligned with the same axis, say, $Z$.
  One can attempt to introduce magnetic anisotropy as a phenomenological
  adjustable parameter, but then one would obtain either a non-symmetric
  magnetic tensor and hence non-reciprocity, which is not physically present in
  the system, or otherwise a tensor whose principal axes do not coincide with
  the crystallographic axes (e.g., for the triangular lattice), or else a
  tensor that is trivially proportional to the identity (e.g., for the square
  lattice). Therefore, anisotropic effective magnetic tensor is also not a
  possibility.}

\bibitem[{\citenamefont{Li et~al.}(2006)\citenamefont{Li, Holt, and
  Efros}}]{li_06_1}
\bibinfo{author}{\bibfnamefont{C.~Y.} \bibnamefont{Li}},
  \bibinfo{author}{\bibfnamefont{J.~M.} \bibnamefont{Holt}}, \bibnamefont{and}
  \bibinfo{author}{\bibfnamefont{A.~L.} \bibnamefont{Efros}},
  \bibinfo{journal}{J. Opt. Soc. Am. B} \textbf{\bibinfo{volume}{23}},
  \bibinfo{pages}{963} (\bibinfo{year}{2006}).

\bibitem[{\citenamefont{Menzel et~al.}(2008)\citenamefont{Menzel, Rockstuhl,
  Paul, Lederer, and Pertsch}}]{menzel_08_1}
\bibinfo{author}{\bibfnamefont{C.}~\bibnamefont{Menzel}},
  \bibinfo{author}{\bibfnamefont{C.}~\bibnamefont{Rockstuhl}},
  \bibinfo{author}{\bibfnamefont{T.}~\bibnamefont{Paul}},
  \bibinfo{author}{\bibfnamefont{F.}~\bibnamefont{Lederer}}, \bibnamefont{and}
  \bibinfo{author}{\bibfnamefont{T.}~\bibnamefont{Pertsch}},
  \bibinfo{journal}{Phys. Rev. B} \textbf{\bibinfo{volume}{77}},
  \bibinfo{pages}{195328} (\bibinfo{year}{2008}).

\bibitem[{\citenamefont{Menzel et~al.}(2010)\citenamefont{Menzel, Rockstuhl,
  Iliew, Lederer, Andryieuski, Malureanu, and Lavrinenko}}]{menzel_10_2}
\bibinfo{author}{\bibfnamefont{C.}~\bibnamefont{Menzel}},
  \bibinfo{author}{\bibfnamefont{C.}~\bibnamefont{Rockstuhl}},
  \bibinfo{author}{\bibfnamefont{R.}~\bibnamefont{Iliew}},
  \bibinfo{author}{\bibfnamefont{F.}~\bibnamefont{Lederer}},
  \bibinfo{author}{\bibfnamefont{A.}~\bibnamefont{Andryieuski}},
  \bibinfo{author}{\bibfnamefont{R.}~\bibnamefont{Malureanu}},
  \bibnamefont{and} \bibinfo{author}{\bibfnamefont{A.~V.}
  \bibnamefont{Lavrinenko}}, \bibinfo{journal}{Phys. Rev. B}
  \textbf{\bibinfo{volume}{81}}, \bibinfo{pages}{195123}
  (\bibinfo{year}{2010}).

\bibitem[{fn4()}]{fn4}
\bibinfo{note}{In a more general geometry, isotropy is not required for
  homogenizability. Thus, for example, a three-dimensional lattice of general
  parallelepipeds can be homogenizable and characterized by the tensor
  $\epsilon_{\rm eff}$ whose all three principal values are
  different~\cite{markel_12_1}.}

\bibitem[{\citenamefont{Stout et~al.}(2006)\citenamefont{Stout, Neviere, and
  Popov}}]{stout_06_1}
\bibinfo{author}{\bibfnamefont{B.}~\bibnamefont{Stout}},
  \bibinfo{author}{\bibfnamefont{M.}~\bibnamefont{Neviere}}, \bibnamefont{and}
  \bibinfo{author}{\bibfnamefont{E.}~\bibnamefont{Popov}}, \bibinfo{journal}{J.
  Opt. Soc. Am. A} \textbf{\bibinfo{volume}{23}}, \bibinfo{pages}{1111}
  (\bibinfo{year}{2006}).

\bibitem[{\citenamefont{Stout et~al.}(2007)\citenamefont{Stout, Neviere, and
  Popov}}]{stout_07_1}
\bibinfo{author}{\bibfnamefont{B.}~\bibnamefont{Stout}},
  \bibinfo{author}{\bibfnamefont{M.}~\bibnamefont{Neviere}}, \bibnamefont{and}
  \bibinfo{author}{\bibfnamefont{E.}~\bibnamefont{Popov}}, \bibinfo{journal}{J.
  Opt. Soc. Am. A} \textbf{\bibinfo{volume}{24}}, \bibinfo{pages}{1120}
  (\bibinfo{year}{2007}).

\bibitem[{\citenamefont{Agranovich and Ginzburg}(1966)}]{agranovich_book_66}
\bibinfo{author}{\bibfnamefont{V.}~\bibnamefont{Agranovich}} \bibnamefont{and}
  \bibinfo{author}{\bibfnamefont{V.}~\bibnamefont{Ginzburg}},
  \emph{\bibinfo{title}{Spatial Dispersion in Crystal Optics and the Theory of
  Excitons}} (\bibinfo{publisher}{Wiley-Interscience}, \bibinfo{address}{New
  York}, \bibinfo{year}{1966}).

\bibitem[{\citenamefont{Landau and Lifshitz}(1984)}]{landau_ess_84}
\bibinfo{author}{\bibfnamefont{L.~D.} \bibnamefont{Landau}} \bibnamefont{and}
  \bibinfo{author}{\bibfnamefont{L.~P.} \bibnamefont{Lifshitz}},
  \emph{\bibinfo{title}{{Electrodynamics of Continuous Media}}}
  (\bibinfo{publisher}{Pergamon Press}, \bibinfo{address}{Oxford},
  \bibinfo{year}{1984}).

\bibitem[{\citenamefont{Agranovich and Gartstein}(2006)}]{agranovich_06_2}
\bibinfo{author}{\bibfnamefont{V.~M.} \bibnamefont{Agranovich}}
  \bibnamefont{and} \bibinfo{author}{\bibfnamefont{Y.~N.}
  \bibnamefont{Gartstein}}, \bibinfo{journal}{Phys. Usp.}
  \textbf{\bibinfo{volume}{49}}, \bibinfo{pages}{1029} (\bibinfo{year}{2006}).

\bibitem[{\citenamefont{Agranovich and Gartstein}(2009)}]{agranovich_09_1}
\bibinfo{author}{\bibfnamefont{V.~M.} \bibnamefont{Agranovich}}
  \bibnamefont{and} \bibinfo{author}{\bibfnamefont{Y.~N.}
  \bibnamefont{Gartstein}}, \bibinfo{journal}{Metamaterials}
  \textbf{\bibinfo{volume}{3}}, \bibinfo{pages}{1} (\bibinfo{year}{2009}).

\bibitem[{\citenamefont{Markel and Tsukerman}(2013)}]{markel_13_1}
\bibinfo{author}{\bibfnamefont{V.~A.} \bibnamefont{Markel}} \bibnamefont{and}
  \bibinfo{author}{\bibfnamefont{I.}~\bibnamefont{Tsukerman}},
  \bibinfo{journal}{Phys. Rev. B} \textbf{\bibinfo{volume}{88}},
  \bibinfo{pages}{125131} (\bibinfo{year}{2013}).

\bibitem[{fn2()}]{fn2}
\bibinfo{note}{Weak nonlocality is usually understood as the regime in which
  Taylor expansion of $\Sigma(\omega, {\bf q})$ to second order in ${\bf q}$ is
  an accurate approximation. For more general nonlinear dependence of
  $\Sigma(\omega, {\bf q})$ on ${\bf q}$, the dispersion equations
  \eqref{PC_Disp_Sigma} or \eqref{PC_Disp_Scal} will have multiple solutions
  that are not obtainable by the operation of folding to first Brillouin zone
  of the lattice.}

\bibitem[{\citenamefont{Luo et~al.}(2002{\natexlab{a}})\citenamefont{Luo,
  Johnson, Joannopoulos, and Pendry}}]{luo_02_1}
\bibinfo{author}{\bibfnamefont{C.}~\bibnamefont{Luo}},
  \bibinfo{author}{\bibfnamefont{S.~G.} \bibnamefont{Johnson}},
  \bibinfo{author}{\bibfnamefont{J.~D.} \bibnamefont{Joannopoulos}},
  \bibnamefont{and} \bibinfo{author}{\bibfnamefont{J.~B.}
  \bibnamefont{Pendry}}, \bibinfo{journal}{Phys. Rev. B}
  \textbf{\bibinfo{volume}{65}}, \bibinfo{pages}{201104}
  (\bibinfo{year}{2002}{\natexlab{a}}).

\bibitem[{\citenamefont{Luo et~al.}(2002{\natexlab{b}})\citenamefont{Luo,
  Johnson, and Joannopoulos}}]{luo_02_2}
\bibinfo{author}{\bibfnamefont{C.}~\bibnamefont{Luo}},
  \bibinfo{author}{\bibfnamefont{S.~G.} \bibnamefont{Johnson}},
  \bibnamefont{and} \bibinfo{author}{\bibfnamefont{J.~D.}
  \bibnamefont{Joannopoulos}}, \bibinfo{journal}{Appl. Phys. Lett.}
  \textbf{\bibinfo{volume}{81}}, \bibinfo{pages}{2352}
  (\bibinfo{year}{2002}{\natexlab{b}}).

\bibitem[{\citenamefont{Craster et~al.}(2011)\citenamefont{Craster, Kaplunov,
  Nolde, and Guenneau}}]{craster_11_1}
\bibinfo{author}{\bibfnamefont{R.~V.} \bibnamefont{Craster}},
  \bibinfo{author}{\bibfnamefont{J.}~\bibnamefont{Kaplunov}},
  \bibinfo{author}{\bibfnamefont{E.}~\bibnamefont{Nolde}}, \bibnamefont{and}
  \bibinfo{author}{\bibfnamefont{S.}~\bibnamefont{Guenneau}},
  \bibinfo{journal}{J. Opt. Soc. Am. A} \textbf{\bibinfo{volume}{28}},
  \bibinfo{pages}{1032} (\bibinfo{year}{2011}).

\bibitem[{\citenamefont{Tsukerman and Markel}(2016)}]{tsukerman_16_1}
\bibinfo{author}{\bibfnamefont{I.}~\bibnamefont{Tsukerman}} \bibnamefont{and}
  \bibinfo{author}{\bibfnamefont{V.~A.} \bibnamefont{Markel}},
  \bibinfo{journal}{Phys. Rev. B} \textbf{\bibinfo{volume}{93}},
  \bibinfo{pages}{024418} (\bibinfo{year}{2016}).

\bibitem[{\citenamefont{Markel and Schotland}(2012)}]{markel_12_1}
\bibinfo{author}{\bibfnamefont{V.~A.} \bibnamefont{Markel}} \bibnamefont{and}
  \bibinfo{author}{\bibfnamefont{J.~C.} \bibnamefont{Schotland}},
  \bibinfo{journal}{Phys. Rev. E} \textbf{\bibinfo{volume}{85}},
  \bibinfo{pages}{066603} (\bibinfo{year}{2012}).

\bibitem[{\citenamefont{Silveirinha}(2007)}]{silveirinha_07_1}
\bibinfo{author}{\bibfnamefont{M.~G.} \bibnamefont{Silveirinha}},
  \bibinfo{journal}{Phys. Rev. B} \textbf{\bibinfo{volume}{75}},
  \bibinfo{pages}{115104} (\bibinfo{year}{2007}).

\bibitem[{\citenamefont{Fietz and Shvets}(2010)}]{fietz_10_1}
\bibinfo{author}{\bibfnamefont{C.}~\bibnamefont{Fietz}} \bibnamefont{and}
  \bibinfo{author}{\bibfnamefont{G.}~\bibnamefont{Shvets}},
  \bibinfo{journal}{Physica B} \textbf{\bibinfo{volume}{405}},
  \bibinfo{pages}{2930} (\bibinfo{year}{2010}).

\bibitem[{\citenamefont{Alu}(2011)}]{alu_11_2}
\bibinfo{author}{\bibfnamefont{A.}~\bibnamefont{Alu}}, \bibinfo{journal}{Phys.
  Rev. B} \textbf{\bibinfo{volume}{84}}, \bibinfo{pages}{075153}
  (\bibinfo{year}{2011}).

\bibitem[{\citenamefont{Markel}(2010)}]{markel_10_2}
\bibinfo{author}{\bibfnamefont{V.~A.} \bibnamefont{Markel}},
  \bibinfo{journal}{J. Phys.: Condens. Matter} \textbf{\bibinfo{volume}{22}},
  \bibinfo{pages}{485401} (\bibinfo{year}{2010}).

\bibitem[{fn3()}]{fn3}
\bibinfo{note}{These assumptions can hold in both 2D and 3D PCs. However, in
  the 2D geometry, even stronger conditions are satisfied: (i) not only ${\bf
  E}_0$ but the total electric field ${\bf E}(x,y) = \hat{\bf z} E(x,y)$ is
  collinear with the $Z$-axis and (ii) the electric field $E(x,y)$ is
  independent of $z$. The latter two properties do not generally hold in
  three-dimensional PCs.}

\bibitem[{\citenamefont{Joannopoulos et~al.}(2008)\citenamefont{Joannopoulos,
  Meade, and Winn}}]{joannopoulos_book_08}
\bibinfo{author}{\bibfnamefont{J.~D.} \bibnamefont{Joannopoulos}},
  \bibinfo{author}{\bibfnamefont{R.~D.} \bibnamefont{Meade}}, \bibnamefont{and}
  \bibinfo{author}{\bibfnamefont{J.~N.} \bibnamefont{Winn}},
  \emph{\bibinfo{title}{Photonic Crystals: Molding the Flow of Light}}
  (\bibinfo{publisher}{Princeton University Press},
  \bibinfo{address}{Princeton, N.J.}, \bibinfo{year}{2008}).

\bibitem[{\citenamefont{Tsukerman and \v{C}ajko}(2008)}]{tsukerman_08_2}
\bibinfo{author}{\bibfnamefont{I.}~\bibnamefont{Tsukerman}} \bibnamefont{and}
  \bibinfo{author}{\bibfnamefont{F.}~\bibnamefont{\v{C}ajko}},
  \bibinfo{journal}{IEEE Trans. Magnetics} \textbf{\bibinfo{volume}{44}},
  \bibinfo{pages}{1382} (\bibinfo{year}{2008}).

\bibitem[{\citenamefont{Gajic et~al.}(2005)\citenamefont{Gajic, Meisels,
  Kuchar, and Hingerl}}]{gajic_05_1}
\bibinfo{author}{\bibfnamefont{R.}~\bibnamefont{Gajic}},
  \bibinfo{author}{\bibfnamefont{R.}~\bibnamefont{Meisels}},
  \bibinfo{author}{\bibfnamefont{F.}~\bibnamefont{Kuchar}}, \bibnamefont{and}
  \bibinfo{author}{\bibfnamefont{K.}~\bibnamefont{Hingerl}},
  \bibinfo{journal}{Opt. Expr.} \textbf{\bibinfo{volume}{13}},
  \bibinfo{pages}{8596} (\bibinfo{year}{2005}).

\bibitem[{\citenamefont{Pei and Huang}(2012)}]{pei_12_1}
\bibinfo{author}{\bibfnamefont{T.-H.} \bibnamefont{Pei}} \bibnamefont{and}
  \bibinfo{author}{\bibfnamefont{Y.-T.} \bibnamefont{Huang}},
  \bibinfo{journal}{J. Opt. Soc. Am. B} \textbf{\bibinfo{volume}{29}},
  \bibinfo{pages}{2334} (\bibinfo{year}{2012}).

\bibitem[{\citenamefont{Tsukerman and Markel}(2014)}]{tsukerman_14_1}
\bibinfo{author}{\bibfnamefont{I.}~\bibnamefont{Tsukerman}} \bibnamefont{and}
  \bibinfo{author}{\bibfnamefont{V.~A.} \bibnamefont{Markel}},
  \bibinfo{journal}{Proc. R. Soc. A} \textbf{\bibinfo{volume}{470}},
  \bibinfo{pages}{0245} (\bibinfo{year}{2014}).

\bibitem[{\citenamefont{Xiong et~al.}(2013)\citenamefont{Xiong, Jiang, Markel,
  and Tsukerman}}]{xiong_13_1}
\bibinfo{author}{\bibfnamefont{X.~Y.~Z.} \bibnamefont{Xiong}},
  \bibinfo{author}{\bibfnamefont{L.~J.} \bibnamefont{Jiang}},
  \bibinfo{author}{\bibfnamefont{V.~A.} \bibnamefont{Markel}},
  \bibnamefont{and}
  \bibinfo{author}{\bibfnamefont{I.}~\bibnamefont{Tsukerman}},
  \bibinfo{journal}{Opt. Expr.} \textbf{\bibinfo{volume}{21}},
  \bibinfo{pages}{10412} (\bibinfo{year}{2013}).

\bibitem[{\citenamefont{Markel}(2008{\natexlab{a}})}]{markel_08_1}
\bibinfo{author}{\bibfnamefont{V.~A.} \bibnamefont{Markel}},
  \bibinfo{journal}{Opt. Expr.} \textbf{\bibinfo{volume}{16}},
  \bibinfo{pages}{19152} (\bibinfo{year}{2008}{\natexlab{a}}).

\bibitem[{\citenamefont{Markel and Schotland}(2010)}]{markel_10_1}
\bibinfo{author}{\bibfnamefont{V.~A.} \bibnamefont{Markel}} \bibnamefont{and}
  \bibinfo{author}{\bibfnamefont{J.~C.} \bibnamefont{Schotland}},
  \bibinfo{journal}{J. Opt.} \textbf{\bibinfo{volume}{12}},
  \bibinfo{pages}{015104} (\bibinfo{year}{2010}).

\bibitem[{\citenamefont{Markel}(2008{\natexlab{b}})}]{markel_08_2}
\bibinfo{author}{\bibfnamefont{V.~A.} \bibnamefont{Markel}},
  \bibinfo{journal}{Phys. Rev. E} \textbf{\bibinfo{volume}{78}},
  \bibinfo{pages}{026608} (\bibinfo{year}{2008}{\natexlab{b}}).

\bibitem[{\citenamefont{Tsukerman}(2008)}]{tsukerman_08_1}
\bibinfo{author}{\bibfnamefont{I.}~\bibnamefont{Tsukerman}},
  \bibinfo{journal}{J. Opt. Soc. Am. B} \textbf{\bibinfo{volume}{25}},
  \bibinfo{pages}{927} (\bibinfo{year}{2008}).

\end{thebibliography}

\appendix

\section{Dispersion equation in a basis-independent form}
\label{app:A}
Let the lattice-periodic permittivity of the medium $\tilde{\epsilon}({\bf r})$ be a bounded, lattice periodic function, possibly a constant. A Bloch mode solution to the scalar wave equation \eqref{wave_eq} can be written in the form
\begin{equation*}
E({\bf r}) =  \tilde{E}({\bf r}) \exp(i {\bf q} \cdot {\bf r}) \ ,
\end{equation*}
where ${\bf q}$ is the Bloch wave vector and $\tilde{E}({\bf r}) \neq 0$ is a nonzero, twice-differentiable, lattice-periodic function, which satisfies the equation
\begin{equation}
\label{a:LF}
{\mathcal L}(\omega,{\bf q}) \tilde{E}({\bf r}) = 0
\end{equation}
with the differential operator ${\mathcal L}(\omega,{\bf q})$ given by
\begin{equation*}
{\mathcal L}(\omega,{\bf q}) = (\nabla + i {\bf q})^2 + k^2 \tilde{\epsilon}({\bf r}) \ .
\end{equation*}
\noindent
Recall that $k = \omega/c$; hence the dependence of ${\mathcal L}$ on $\omega$. The other reason for this dependence is frequency dispersion (dependence of $\tilde{\epsilon}$ on $\omega$), which is not indicated here explicitly but is taken into account. The requirement that \eqref{a:LF} has a nontrivial solution $\tilde{E}({\bf r}) \neq 0$ defines the dispersion equation $f(\omega, {\bf q}) = 0$. We will assume without proof the following statement: if for some pair $(\omega, {\bf q})$ \eqref{a:LF} has a nontrivial solution, then this solution is unique up to multiplication by a constant. This is not generally true in 3D where different polarization states can correspond to the same pair $(\omega, {\bf q})$.

We will refer to the pairs $(\omega, {\bf q})$ for which \eqref{a:LF} has a nontrivial solution as to the solutions to the dispersion equation. Here the variables $\omega$ and ${\bf q}$ are not restricted and can take general complex values. However, when studying stationary processes, one is typically interested only in solutions with $\omega>0$. We also note that, if $(\omega,{\bf q})$ is a solution, then $(\omega,{\bf q} + {\bf g})$ is also a solution, where ${\bf g}$ is any reciprocal lattice vector. This mathematical property of Bloch waves gives rise to ``folding'' of the solutions to the dispersion equation. 

It is a nontrivial mathematical question how to distinguish between the solutions that occur due to folding from those that occur due to multiple scattering (interaction). In particular, it is quite plausible that solutions $(\omega,{\bf q})$ and $(\omega,{\bf q}^\prime)$ of different physical origin can co-exist at the same frequency $\omega$ (or at two very close frequencies). In this Appendix, we present an approach to mathematical classification of these two physically-different solutions.

As was done in Ref.~\onlinecite{tsukerman_08_1}, we decompose $\tilde{E}({\bf r})$ and $\tilde{\epsilon}({\bf r})$ into the constant and zero-mean components according to
%
\begin{align*}
& \tilde{E}({\bf r}) = E_0 + F({\bf r}) \ , \ \ \langle F({\bf r}) \rangle_{\mathbb{C}} = 0 \ , \\
& \tilde{\epsilon}({\bf r}) = \epsilon_0 + \eta({\bf r}) \ , \ \ \langle \eta({\bf r}) \rangle_{\mathbb{C}} = 0 \ .
\end{align*}
%
\noindent
where $\langle \ldots \rangle_{\mathbb{C}}$ indicates averaging over the elementary cell $\mathbb{C}$. We note that $E_0$ is the amplitude of the fundamental harmonic of the Bloch wave. Upon substitution of the above decomposition into \eqref{a:LF}, we obtain
\begin{equation}
\label{a:L_F0_Ft}
\left[ k^2 (\epsilon_0 +\eta({\bf r})) - q^2 \right] E_0 + {\mathcal L}(\omega,{\bf q}) F({\bf r}) = 0 \ .
\end{equation}
\noindent
We now introduce the averaging operator ${\mathcal O}$ and the operator ${\mathcal P} = 1 - {\mathcal O}$ as projections onto the complementary subspaces of constant and zero-mean functions so that ${\mathcal O} \tilde{E}({\bf r}) = E_0$, ${\mathcal P} \tilde{E}({\bf r}) = F({\bf r})$ and ${\mathcal O} \tilde{\epsilon}({\bf r}) = \epsilon_0$, ${\mathcal P} \tilde{\epsilon}({\bf r}) = \eta({\bf r})$. By acting on \eqref{a:L_F0_Ft} with ${\mathcal O}$ and ${\mathcal P}$ from the left, we obtain the following two equations
\begin{subequations}
\label{a:proj}
\begin{align}
\label{a:proj_O}
& \left( k^2 \epsilon_0 - q^2 \right) E_0 + k^2 \left \langle \eta({\bf r}) F ({\bf r}) \right\rangle_{\mathbb C} = 0  \ , \\
\label{a:proj_P}
& \eta({\bf r}) E_0 + {\mathcal W}(\omega,{\bf q}) F({\bf r}) = 0  \ .
\end{align}
\end{subequations}
\noindent
In the first equation above, we have used the equality ${\mathcal O} {\mathcal L} F({\bf r}) = k^2 {\mathcal O} \eta({\bf r}) F({\bf r}) = k^2 \left\langle \eta({\bf r}) F({\bf r}) \right\rangle_{\mathbb C}$, which can be proved by using cell-periodicity of $F({\bf r})$ and integral theorems. In the second equation, we have used $({\mathcal P}{\mathcal L}) F({\bf r}) = ({\mathcal P}{\mathcal L} {\mathcal P}) F({\bf r})$ and defined the operator 
\begin{equation*}
{\mathcal W}(\omega, {\bf q}) = k^{-2}{\mathcal P} {\mathcal L}(\omega, {\bf q}) {\mathcal P} \ .
\end{equation*}
\noindent
We can now consider the following two different kinds of solutions to \eqref{a:proj}:

\begin{enumerate}

\item 
Solutions with $E_0 \neq 0$. Bloch waves $E({\bf r})$ with nonzero fundamental harmonic $E_0\neq 0$ can satisfy the wave equation \eqref{wave_eq} for a given pair $(\omega, {\bf q})$ only if the equation ${\mathcal W}(\omega, {\bf q}) \phi = \eta$ has a nonzero solution. According to the assumption of uniqueness of $\tilde{E}$ stated above, this solution is unique if it exists. We will say that $\phi$ is given in this case by the inverse of ${\mathcal W}$, viz, $\phi = {\mathcal W}^{-1} \eta$. Then it follows from \eqref{a:proj_P} that $F = - E_0 {\mathcal W}^{-1} \eta$. Substituting this expression into \eqref{a:proj_O}, we arrive at the dispersion relation \eqref{PC_Disp_Scal} in which $\Sigma(\omega, {\bf q})$ is given by the following basis-independent expression:
\begin{equation}
\label{a:Sigma_def}
\Sigma(\omega, {\bf q}) = \epsilon_0 - \left\langle \eta({\bf r}) {\mathcal W}^{-1}(\omega, {\bf q}) \eta ({\bf r}) \right\rangle_{\mathbb C} \ .
\end{equation}

\item 
Solutions with $E_0=0$. There can exist another class of solutions, that is, Bloch waves with a zero fundamental harmonic. For such solutions to exist, the equation ${\mathcal W}(\omega, {\bf q}) \phi = 0$ must have a nontrivial solution such that $\left\langle \eta ({\bf r}) \phi ({\bf r}) \right\rangle_{\mathbb C} = 0$. In this case, ${\mathcal W}(\omega, {\bf q})$ is singular. Solutions of this kind exist in the case of a homogeneous medium that is artificially discretized on an arbitrary lattice (see below). However, solutions with $E_0 = 0$ (or, in practice, with $E_0$ in some sense very small) can also exist in heterogeneous periodic media with $\eta \neq 0$.

\end{enumerate}

We will say that solutions to the dispersion equation of the first kind generate ``true'' $\Gamma$-point frequencies that result from multiple scattering in inhomogeneous photonic crystals. Solutions of the second kind are due to purely geometrical folding and generate spurious $\Gamma$-point frequencies that do not satisfy the condition \eqref{Gamma_def_gen_b}. 

This classification can be illustrated by considering the special case of a homogeneous medium with $\epsilon({\bf r}) = \epsilon_0$ that is artificially discretized on an arbitrary lattice of a finite pitch. In this case, it is easy to see that ${\mathcal W}(\omega, {\bf q})$ is singular if $({\bf q} + {\bf g})^2 = k^2 \epsilon_0$ where ${\bf g} \neq 0$ is any {\em nonzero} reciprocal lattice vector, and invertible otherwise. Therefore, the dispersion equation for the solutions of the first kind is of the form 
\begin{equation*}
q^2 = k^2 \epsilon_0 \ , \ \  \Sigma(\omega, {\bf q}) = \epsilon_0 \ \ {\rm (first \ kind)} \ .
\end{equation*}
\noindent
As could be expected, the corresponding lattice-periodic function $\tilde{E}({\bf r}) = E_0$ is just a constant. Thus, the fundamental harmonic is dominating in this solution.

Solutions of the second kind are of the form
\begin{equation}
\label{a:de_hm}
({\bf q} + {\bf g})^2 = k^2\epsilon_0 \ \ {\rm for} \ {\bf g}\neq 0 \ \ {\rm (second\ kind)} \ .
\end{equation}
\noindent
For any pair $(\omega, {\bf q})$ satisfying the above condition, ${\mathcal W}(\omega, {\bf q})$ is singular and we have $\langle \tilde{E} \rangle_{\mathbb C} = 0$. This lattice-periodic function is dominated by higher-order harmonics. 

We can investigate the spurious $\Gamma$-point frequencies generated by the solution of the second kind as follows. If ${\rm Im}(\epsilon_0) = 0$, equation \eqref{a:de_hm} is satisfied by a pair $(\omega_{\bf p}, 0)$ where $\omega_{\bf p}^2 = (cp)^2/\epsilon_0$ and ${\bf p} \neq 0$ is any nonzero reciprocal lattice vector. The isofrequency line at $\omega = \omega_{\bf p}$ is then obtained from the equation 
\begin{equation}
\label{a:iso_hmg}
({\bf q} + {\bf g})^2 = p^2>0 \ \ {\rm for\ any\ } {\bf g} \neq 0 \ . 
\end{equation}
\noindent
Consider for simplicity a square lattice and let $\epsilon_0>0$ and ${\bf p} = (2\pi/a)(0,1)$. The corresponding quantity $\omega_{\bf p} = 2 \pi c / a\sqrt{\epsilon_0}$ is the first spurious $\Gamma$-point frequency out of the infinite sequence. In principle, equation \eqref{a:iso_hmg} defines an infinite number of curves but, at the first $\Gamma$-point frequency, most of these curves coincide. To obtain the complete solution, we can start by taking ${\bf g} = {\bf p}$. It can be seen that \eqref{a:iso_hmg} defines in this case a circular arc in the $(q_x,q_y)$ plane, and that this arc crosses the origin. On a square lattice, there are four reciprocal vectors of the same length $2\pi/a$ and pointing in the directions $(0,\pm 1)$ and $(\pm 1,0)$. As a result, the isofrequency line contains four circular arcs intersecting at the origin (six arcs in the case of a triangular lattice) as is shown in Fig.~\ref{fig:IF_Hmg}(a). The arcs are truncated and reflected at the edge of the FBZ. These additional ``reflected'' segments of the isofrequency line can be obtained by considering additional vectors ${\bf p}$ in \eqref{a:iso_hmg}.

The above result is quite trivial and we could have obtained it without using the mathematical formalism of this Appendix. However, the derivation is useful to show that the formalism developed herein is consistent with the limit of zero contrast. More importantly, it provides us with a clue how to treat the case when $E_0$ is small but nonzero. Let $\gamma = \Vert  F / E_0 \Vert_2$ where $\Vert \cdot \Vert_2$ is the $L_2$ norm. We believe that it is physically meaningful to consider the phase and group velocities of a Bloch wave only if $\gamma$ is sufficiently small so that the fundamental harmonic of the Bloch wave is in some sense dominant. This condition holds if the smallest singular value of ${\mathcal W}$ is sufficiently far from zero. If this is not so, then $\gamma$ can be very large or even diverge. In this case, introducing the characteristics of a plane wave such as the phase and group velocities is devoid of physical meaning, even if this can be done formally by considering the dependence $q(\omega)$. 

We therefore conjecture that there are two fundamentally different regimes of propagation, $\gamma \ll 1$ and $\gamma \gg 1$, and in the second regime the group velocity computed as, say, $\nabla_{\bf q} \omega$ does not correspond to any physically measurable quantity and should not be invested with any particular interpretation. It is not clear, though, how to treat the borderline case $\gamma \sim 1$; perhaps, it can only be investigated numerically.

It should be noted that the above discussion is applicable to the case of s-polarization when the scalar electric field is smooth. For the p-polarization, the electric field can jump at the discontinuities of $\tilde{\epsilon}({\bf r})$, and a more careful analysis is required. We also note that the condition on $\gamma$ stated above is closely related to the concept of the smooth field introduced in Ref.~\onlinecite{markel_12_1}. 

\section{Dispersion equation and perturbation theory in the basis of plane waves}
\label{app:B}

In this Appendix, we derive equation \eqref{PC_Disp_Scal} from \eqref{main}, define the function $\Sigma(\omega, {\bf q})$ algebraically using the basis of plane waves and then obtain the expansion of $\Sigma(\omega, {\bf q})$ in powers of the Cartesian components of ${\bf q}$ for $\omega$ in the vicinity of one of the $\Gamma$-point frequencies $\omega_n$ [as defined by \eqref{Gamma_def_gen}]. We note that a basis-independent definition of $\Sigma(\omega, {\bf q})$ is given in Appendix~\ref{app:A} by equation \eqref{a:Sigma_def}. In order to derive \eqref{a:Sigma_def}, we have assumed that the fundamental Bloch harmonic $E_0$ is non-zero. We will use this assumption in this Appendix as well. We are therefore restricting ourselves to the solutions of the first kind according to the terminology of Appendix~\ref{app:A}.

As was done by us previously~\cite{markel_12_1,markel_13_1}, we consider the equation \eqref{main} for the cases ${\bf g} = 0$ and ${\bf g} \neq 0$ separately. Since we assume that $E_0 \neq 0$, we can scale all coefficients $E_{\bf g}$ by $E_0$. Let $e_{\bf g} = E_{\bf g}/E_0$, so that $e_0 = 1$. First, we take ${\bf g}=0$ in \eqref{main}. This results in 
\begin{eqnarray}
\label{b:1}
&q^2& = k^2 \left[ \epsilon_h + \rho\chi\sum_{\bf p} M(-{\bf p}) e_{\bf p} \right] \nonumber \\
    &=& k^2 \left[ \epsilon_h + \rho\chi M(0)e_0 + \rho\chi\sum_{{\bf p}\neq 0} M(-{\bf p}) e_{\bf p} \right] \ .
\end{eqnarray}
\noindent
Since $M(0)e_0 = 1$ and $\chi = \epsilon_i - \epsilon_h$, we see that \eqref{b:1} is equivalent to \eqref{PC_Disp_Scal} in which
\begin{equation}
\label{Sigma_def}
\Sigma(\omega, {\bf q}) = \epsilon_0 + \rho\chi \sum_{{\bf g} \neq 0} M(-{\bf g}) e_{\bf g} 
\end{equation}
\noindent
and $\epsilon_0 = \rho \epsilon_i + (1-\rho) \epsilon_h$ is the cell-averaged permittivity of the PC, the same quantity as the zeroth component in the Fourier expansion \eqref{eps_exp}.

We now must show that the expression in the right-hand side of \eqref{Sigma_def} is a well-defined quantity and a function of the two arguments $\omega$ and ${\bf q}$. To this end we note that this expression contains the relative amplitudes $e_{\bf g} = E_{\bf g}/E_0$ with ${\bf g} \neq 0$. Therefore, to compute $\Sigma(\omega, {\bf q})$ algebraically, we must consider \eqref{main} for ${\bf g} \neq 0$. This yields the following set of equations
\begin{align}
\label{g=/=0}
[({\bf q} + &{\bf g} )^2 - k^2\epsilon_h ]e_{\bf g} = \rho\chi k^2 \nonumber \\ &\times \left[ M({\bf g}) + \sum_{{\bf p} \neq 0} M({\bf g} - {\bf p}) e_{\bf p} \right] \ , \ \ {\bf g} \neq 0 \ .
\end{align}
\noindent
It is important to note that \eqref{g=/=0} is a closed (albeit an infinite) set of equations with a nonzero free term $\rho \chi k^2 M({\bf g})$. Consequently, \eqref{g=/=0} is not an eigenproblem but rather a linear set of equations that can be, in principle, solved by matrix inversion. This proves that \eqref{Sigma_def} is a mathematically-consistent definition of the function $\Sigma(\omega, {\bf q})$.

To complete the algebraic definition of $\Sigma(\omega, {\bf q})$, we can proceed as follows. Define the matrices and vectors:
%
\begin{align*}
& D_{{\bf g}{\bf p}} = \left(g^2 - k^2\epsilon_h \right) \delta_{{\bf g}{\bf p}} \ , \\
& Q_{{\bf g}{\bf p}} = \left( q^2 + 2 {\bf q} \cdot {\bf g} \right) \delta_{{\bf g}{\bf p}} \  , \\
& M_{{\bf g}{\bf p}} = M({\bf g} - {\bf p}) \ , \\
& b_{\bf g}          = M({\bf g}) \ .
\end{align*}
%
\noindent
Here all indexes are restricted to ${\bf g} \neq 0$ and ${\bf p}\neq 0$. Also, only the matrix $Q$ depends on ${\bf q}$. We have separated the diagonal term $Q$ from the ${\bf q}$-independent diagonal term $D$ because we are interested in building a perturbation theory in Cartesian components of ${\bf q}$. Also note the normalization rule $\langle b \vert b \rangle = 1/\rho-1$. We can now re-write \eqref{g=/=0} as follows:
\begin{equation}
\label{g=/=0_oper}
(Q + D - \rho\chi k^2 M)\vert e \rangle = \rho \chi k^2 \vert b \rangle \ .
\end{equation}
From this we find that
\begin{equation*}
\Sigma(\omega, {\bf q}) = \epsilon_0 + (\rho \chi k)^2 \langle b \vert (D + Q - \rho\chi k^2 M)^{-1} \vert b \rangle \ .
\end{equation*}
\noindent
This expression gives a closed-form algebraic definition of $\Sigma(\omega, {\bf q})$.

We now proceed with building the perturbation theory. We assume that (i) a higher-order $\Gamma$-point frequency $\omega_n$ exists according to the definition \eqref{Gamma_def_gen} and (ii) $\omega$ is in the vicinity and just below $\omega_n$ (in the pass-band). Under this condition, $\Sigma(\omega, 0)$ is small but nonzero and positive. Otherwise, the frequency $\omega$ would be is inside a bandgap. We can define the $T$-matrix of the problem as follows:
\begin{equation}
\label{T_def}
T = \left( D - \rho\chi k^2 M \right)^{-1} \ .
\end{equation}
\noindent
Note that $T$ is the matrix that we need in order to compute $\Sigma(\omega, 0)$. Indeed, for ${\bf q}=0$, we also have $Q=0$ and 
\begin{eqnarray*}
 \Sigma(\omega, 0) &=& \epsilon_0 + (\rho\chi k)^2 \langle b \vert T \vert b \rangle \nonumber \\
                   &=& \epsilon_0 + (\rho\chi k)^2 \sum_{{\bf g}_1,{\bf g}_2 \neq 0} M(-{\bf g}_1) T_{{\bf
    g}_1 {\bf g}_2} M({\bf g}_2) \ .
\end{eqnarray*}
\noindent
Let us assume that $T = T(\omega)$ has been computed at the working frequency $\omega$ by inverting matrix in the right-hand side of \eqref{T_def}. Computationally, this requires truncation of the basis and one matrix inversion operation. Now, the equation we intend to iterate is of the form
\begin{equation*}
\vert e \rangle = \rho\chi k^2 T \vert b \rangle - TQ \vert e \rangle \ ,
\end{equation*}
\noindent
which follows directly from \eqref{g=/=0_oper} and \eqref{T_def}. The formal power series solution is of the form
\begin{equation*}
\vert e \rangle = \rho\chi k^2 \sum_{n=0}^\infty (-TQ)^n T \vert b \rangle \ ,
\end{equation*}
\noindent
and for $\Sigma$, we have
%
\begin{align*}
\Sigma &= \epsilon_0 + (\rho\chi k)^2 \sum_{n=0}^\infty \langle b \vert
(-TQ)^n T \vert b \rangle \\
&= \epsilon_0 + (\rho\chi k)^2 \sum_{n=0}^\infty \sigma_n \ ,
\end{align*}
%
\noindent
where 
\begin{equation*}
\sigma_n = (-1)^n \langle b \vert (TQ)^nT \vert b \rangle \ . 
\end{equation*}
\noindent
Note that the order in $Q$ is not the same as the order in ${\bf q}$. Also note that terms of the form 
\begin{equation}
\label{terms_iso}
(-1)^j q^{2j} \langle b \vert T^{j+1} \vert b \rangle
\end{equation}
\noindent
are generated in all orders of the expansion. These terms are, of course, perfectly isotropic. However, starting from fourth order, more complicated terms appear in the expansion.

In the formulas of this Appendix, we will use the following notations:
\begin{itemize}
\item[(i)] The symbol $\circ$ denotes direct (Hadamard) product of two matrices, e.g., $(A\circ B)_{ij} = A_{ij} B_{ij}$. 

\item[(ii)] The matrices $G$, $G_{xx}$, etc., are defined in terms of the reciprocal lattice vectors as follows:
%
\begin{align*}
& (G_{\alpha\beta})_{{\bf g}{\bf p}} =  g_\alpha p_\beta \ , \ \ \alpha,\beta=x,y \ ; \\
& G_{{\bf g}{\bf p}} =  \sum_\alpha G_{\alpha\alpha} =  {\bf g} \cdot {\bf p} \ .
\end{align*}
%
\item[(iii)] In all expressions shown below we imply that, for example, $q^6 = (q_x^2 + q_y^2)^3$, etc.
\end{itemize}

\paragraph{Zeroth order.} We start with zero order ($n=0$), which is rather trivial:
\begin{equation*}
\sigma_0 = \langle b \vert T \vert b \rangle = \sum_{{\bf g}_1, {\bf g}_2} M(-{\bf g}_1) T_{{\bf g}_1 {\bf g}_2} M({\bf g}_2) \ .
\end{equation*}
\noindent
Thus, $\sigma_0$ is a ${\bf q}$-independent constant. Obviously, $\Sigma(\omega,0) = \epsilon_0 + (\rho\chi k)^2 \sigma_0(k)$, where we have indicated the dependence of $\sigma_0$ on $k$ explicitly.

\paragraph{First order.} Here we have
\begin{align*}
\sigma_1  = - \sum_{{\bf g}_1,{\bf g}_2, {\bf g}} & M(-{\bf g}_1) T_{{\bf g}_1 {\bf g}} \left(q^2 + 2{\bf q} \cdot {\bf g} \right) \nonumber \\
         & \times T_{{\bf g} {\bf g}_2}  M({\bf g}_2) \ .                                     
\end{align*}
\noindent
The contribution of the second term in the brackets is zero due to the symmetry [$\sigma_n(-{\bf q}) = \sigma_n({\bf q})$]. Therefore,
\begin{align*}
\sigma_1 &= - q^2 \sum_{{\bf g}_1,{\bf g}_2}  M(-{\bf g}_1) T^2_{{\bf g}_1 {\bf g}_2} M(-{\bf g}_2) \nonumber \\
         &= - q^2 \langle b \vert T^2 \vert b \rangle \ .
\end{align*}

\paragraph{Second order.} Here we have
\begin{align*}
\sigma_2  & = \sum_{\{{\bf g}\}} M(-{\bf g}_1) T_{{\bf g}_1 {\bf g}_2} T_{{\bf g}_2 {\bf g}_3} T_{{\bf g}_3 {\bf g}_4} M({\bf g}_4) \nonumber \\
& \times \left[q^2 + 2 ({\bf q} \cdot {\bf g}_2) \right] 
         \left[q^2 + 2 ({\bf q} \cdot {\bf g}_3) \right] \ ,
\end{align*}
\noindent
where $\sum_{\{{\bf g}\}}$ denotes summation over all relevant indexes. We now expand the product of the square brackets and notice that the terms linear in ${\bf q}$ sum to zero for the reason of inversion symmetry. The two terms that produce nonzero result upon summation are 
\begin{equation*}
q^4 \ \ \  {\rm and} \ \ \ 4 ({\bf q}\cdot {\bf g}_2) ( {\bf q} \cdot {\bf g}_3 ) \ .
\end{equation*}
\noindent
The first of these generates the result of the form \eqref{terms_iso} and the second term requires some additional consideration. We can write
\begin{align*}
({\bf q} \cdot {\bf g}_2) & ( {\bf q} \cdot {\bf g}_3 )
= (q_x g_{2x} + q_y g_{2y}) (q_x g_{3x} + q_y g_{3y}) \\
&= q_x^2 g_{2x}g_{3x} + q_y^2 g_{2y} g_{3y} + q_x q_y (g_{2x} g_{3y} + g_{2y} g_{3x}) \ .
\end{align*}
\noindent
The term proportional to $q_x q_y$ sums to zero due the inversion symmetry. Moreover, from the symmetry properties of both triangular and square lattices, we find that summation of the coefficients in front of $q_x^2$ and $q_y^2$ must yield the same
result. Indeed, if this were not so, we would have obtained a term of the form $\beta_x q_x^2 + \beta_y q_y^2$ with $\beta_x \neq \beta_y$, describing an ellipse of unequal semiaxes, which is inconsistent with both $C_4$ and $C_6$ symmetries. Therefore, we can replace the above expression (inside the summation) by
\begin{align*}
\frac{1}{2}(q_x^2 + q_y^2) (g_{2x}g_{3x} + g_{2y} g_{3y}) = \frac{1}{2} q^2 ({\bf g}_2 \cdot {\bf g}_3) \ .
\end{align*}
\noindent
Collecting everything together, we find that
\begin{align*}
\sigma_2 = q^4 \langle b \vert T^3 \vert b \rangle + 2q^2 \langle b \vert T (T\circ G) T \vert b \rangle \ .
\end{align*}
\noindent
We thus see that all terms generated in the second order are still circularly-symmetric. 

\paragraph{Third order.} Here we have 
\begin{align*}
\sigma_3 &= -\sum_{\{{\bf g}\}}M(-{\bf g}_1) T_{{\bf g}_1 {\bf g}_2} T_{{\bf g}_2 {\bf g}_3} T_{{\bf g}_3 {\bf g}_4} T_{{\bf g}_4 {\bf g}_5} M({\bf g}_5) \nonumber \\
\times & \left[q^2 + 2({\bf q} \cdot {\bf g}_2 ) \right] 
         \left[q^2 + 2({\bf q} \cdot {\bf g}_3 ) \right]
         \left[q^2 + 2({\bf q} \cdot {\bf g}_4 ) \right] \ .
\end{align*}
\noindent
After expanding the brackets and keeping only the terms that do not sum to zero, we find that $\sigma_3 = \sigma_3^{(a)} + \sigma_3^{(b)}$ where
\begin{equation}
\label{sigma_3a}
\sigma_3^{(a)} = -q^6 \langle b \vert T^4 \vert b \rangle
\end{equation}
\noindent
is of the form \eqref{terms_iso} and 
\begin{align*}
& \sigma_3^{(b)} = -4q^2 \sum_{\{{\bf g}\}}M(-{\bf g}_1) T_{{\bf g}_1 {\bf g}_2} T_{{\bf g}_2 {\bf g}_3} T_{{\bf g}_3 {\bf g}_4}
T_{{\bf g}_4 {\bf g}_5} M({\bf g}_5) \nonumber \\
& \times \left[ ({\bf q} \cdot {\bf g}_2 ) ({\bf q} \cdot {\bf g}_3) +
                ({\bf q} \cdot {\bf g}_2 ) ({\bf q} \cdot {\bf g}_4) +
                ({\bf q} \cdot {\bf g}_3 ) ({\bf q} \cdot {\bf g}_4)
                  \right] \ .
\end{align*}
\noindent
We can use the same transformation as was used in the second order to transform the terms of the form $({\bf q} \cdot {\bf
  g}_2) ({\bf q} \cdot {\bf g}_3)$ to the form $(1/2)q^2 ({\bf g}_2 \cdot {\bf g}_3)$. We thus obtain 
\begin{align}
\label{sigma_3b}
\sigma_3^{(b)} &= 2(qh)^4 \Big{[} \langle b \vert T^2 (T\circ G)    T
\vert b \rangle \nonumber \\
 &+ \langle b \vert T   (T^2 \circ G) T  \vert b \rangle + 
   \langle b \vert T   (T\circ G)    T^2 \vert b \rangle \Big{]} \ .
\end{align}
\noindent
The first and last terms in this expression are in fact equal but are written separately for symmetry of expression. Still, all expressions arising to third order in $Q$ are circularly-symmetric.

\paragraph{Fourth order.} In the fourth order, we can write
\begin{equation*}
\sigma_4 = \sigma_4^{(a)} + \sigma_4^{(b)} + \sigma_4^{(c)} \ ,
\end{equation*}
\noindent
where the expressions for $\sigma_4^{(a)}$ and $\sigma_4^{(b)}$ are obtained in the manner very similar to what was done above. Omitting the intermediate steps, we write the final result for these two terms, viz,
\begin{align}
\label{sigma_4a}
\sigma_4^{(a)} = q^8 \langle b \vert T^5 \vert b\rangle
\end{align}
\begin{align}
\label{sigma_4b}
\sigma_4^{(b)} &= 2q^6  \nonumber \\
\big{[}        &\langle b \vert T   (T\circ G)   T^3 \vert b\rangle  
              + \langle b \vert T^3 (T\circ G)   T   \vert b\rangle \nonumber \\
              +&\langle b \vert T   (T^2\circ G) T^2 \vert b\rangle
              + \langle b \vert T^2 (T^2\circ G) T   \vert b\rangle \nonumber \\
              +&\langle b \vert T^3 (T\circ G)   T   \vert b\rangle
              + \langle b \vert T^2 (T\circ G)   T^2 \vert b\rangle
              \big{]} \ .
\end{align}
\noindent
Note that the terms appearing on each line of the above expression are pair-wise equal. The expressions \eqref{sigma_4a} and
\eqref{sigma_4b} could, in fact, be anticipated and are directly analogous to the expressions \eqref{sigma_3a} and \eqref{sigma_3b}. A simple diagrammatic technique can be devised to generate similar expressions that appear in the higher orders of the perturbation theory.

However, $\sigma_4^{(c)}$ is a term of a different kind and it is the first term we encounter that does not obey the circular symmetry and is consistent with $C_4$ (but not $C_6$) symmetry. The term is
\begin{align}
\label{sigma_4c_i}
\sigma_4^{(c)} &= 16  \sum_{\{{\bf g}\}}M(-{\bf g}_1) T_{{\bf
    g}_1 {\bf g}_2} \ \ldots \ T_{{\bf g}_5 {\bf g}_6} M({\bf g}_6) \nonumber \\
              &\times ({\bf q} \cdot {\bf g}_2) ({\bf q} \cdot {\bf g}_3) 
({\bf q} \cdot {\bf g}_4) ({\bf q} \cdot {\bf g}_5) \ .
\end{align}
\noindent
We can not use the same trick as was used above to transform the factor 
\begin{align*}
P = ({\bf q} \cdot {\bf g}_2) ({\bf q} \cdot {\bf g}_3) 
({\bf q} \cdot {\bf g}_4) ({\bf q} \cdot {\bf g}_5)
\end{align*}
\noindent
to the form $\beta q^4$ where $\beta$ is a scalar expressible in terms of the dot products ${\bf g}_i \cdot {\bf g}_j$. Indeed, let us write the factor $P$ in terms of Cartesian components of all vectors involved:
\begin{align*}
P &=       (q_x g_{2x} + q_y g_{2y}) (q_x g_{3x} + q_y g_{3y}) \nonumber \\
  &\times  (q_x g_{4x} + q_y g_{4y}) (q_x g_{5x} + q_y g_{5y})
  \nonumber \\
  &= P^\prime + q_x^4 g_{2x}g_{3x}g_{4x}g_{5x} + q_y^4 g_{2y}g_{3y}g_{4y}g_{5y}
  \nonumber \\
+ q_x^2 q_y^2 \big{[} 
  & g_{2x}g_{3x}g_{4y}g_{5y} + g_{2y}g_{3y}g_{4x}g_{5x} \nonumber \\
+ & g_{2x}g_{3y}g_{4x}g_{5y} + g_{2y}g_{3x}g_{4y}g_{5x} \nonumber \\
+ & g_{2x}g_{3y}g_{4y}g_{5x} + g_{2y}g_{3x}g_{4x}g_{5y} \Big{]} \ .
\end{align*}
\noindent
Here $P^\prime$ is the term that sums to zero by symmetry. We can introduce the notations $\Delta=P-P^\prime$ and $\Pi_x$, $\Pi_y$, $\Pi_{xy}$ (definition of the last three quantities will be clear from the next equation) and write
\begin{equation*}
\Delta = \Pi_x q_x^4 + \Pi_y q_y^4 + \Pi_{xy} q_x^2 q_y^2
\end{equation*}
\noindent
Now, if it happens so that
\begin{equation*}
\sum_{\{{\bf g}\}}F[\{{\bf g}\}] \Pi_x = \sum_{\{{\bf g}\}}F[\{{\bf
  g}\}] \Pi_y =  \frac{1}{2}\sum_{\{{\bf g}\}}F[\{{\bf g}\}] \Pi_{xy} \ ,
\end{equation*}
\noindent
where $F[\{{\bf g}\}]$ is the coefficient appearing on the first line of \eqref{sigma_4c_i}, then we would obtain the result $\sigma_4^{(c)} = \beta q^4$. This is what we can expect to happen in triangular lattices with $C_6$ symmetry. However, there is no general or obvious reason why the second equality in the above expression should hold in the case of $C_4$ symmetry, and there are sufficient grounds to believe that it does not. As a result, the function $\Delta({\bf q})$ and the summation result
$\sum_{\{{\bf g}\}}F[\{{\bf g}\}] \Delta({\bf q})$ are not circularly-symmetric. 

We can write the result for $\sigma_4^{(c)}$ in a form similar to that used in lower orders 
if we account for the identity $q_x^4 + q_y^4 = q^4 - 2q_x^2 q_y^2$. Then
\begin{align*}
\sigma_4^{(c)} &= 8 \left[q^4 - 2 (q_x^2q_y^2) \right] \nonumber
\\
&\times \Big{[} \langle b \vert T (T\circ G_x) T (T\circ G_x) T\vert b
\rangle + (G_x \rightarrow G_y) \Big{]} \nonumber \\
& + q_x^2q_y^2 \nonumber \\
\times \Big{[} 
&\langle b \vert T (T\circ G_x) T (T\circ G_y) T\vert b \rangle + (G_x
\leftrightarrow G_y) \nonumber \\
+&\langle b \vert T (T\circ G_{xy}) T (T\circ G_{xy}) T\vert b \rangle + (G_{xy}
\rightarrow G_{yx}) \nonumber \\
+&\langle b \vert T (T\circ G_{xy}) T (T\circ G_{yx}) T\vert b \rangle + (G_{xy}
\leftrightarrow G_{yx}) \Big{]} \ .
\end{align*} 

The terms of the form \eqref{C6} appear only in the sixth order of the perturbation theory. The corresponding coefficients are very complicated and we do not compute them here.

\end{document}